\documentclass[12pt]{iopart}

\usepackage{graphicx,xcolor}
\expandafter\let\csname equation*\endcsname\relax
\expandafter\let\csname endequation*\endcsname\relax
\usepackage{amsmath, amsthm, amssymb}
\usepackage{float}
\usepackage{subfig}
\usepackage{xcolor}
\usepackage{fancyhdr}
\UseRawInputEncoding

\begin{document}
\title{{Minimum-q induced alternation} between infernal modes and EP-driven modes
	in advanced tokamak configurations}
\author{Shiwei Xue}

\address{State Key Laboratory of Advanced Electromagnetic Technology, \\International Joint Research Laboratory of Magnetic Confinement Fusion and Plasma Physics, School of Electrical and Electronic Engineering,
		\\	Huazhong University of Science and Technology, Wuhan, 430074,
		China}

\author{Ping Zhu*}

\address{State Key Laboratory of Advanced Electromagnetic Technology, \\International Joint Research Laboratory of Magnetic Confinement Fusion and Plasma Physics, School of Electrical and Electronic Engineering,
		\\	Huazhong University of Science and Technology, Wuhan, 430074,
		China;
~\\
Department of Nuclear Engineering and Engineering Physics, 
\\University of Wisconsin-Madison, Madison,
Wisconsin, 53706, United States of America}
\ead{zhup@hust.edu.cn}

\author{Haolong Li}

\address{College of Sciences, Tianjin University of Science and Technology, Tianjin 300457, China}

\vspace{10pt}
\begin{indented}
\item[]\today
\end{indented}
\clearpage
\begin{abstract}

For an advanced tokamak configuration
in the presence of energetic particles (EPs), the dominant instability is found to alternate between infernal modes and Alfv\'en eigenmodes with the variation of the minimum safety factor $q_{\min}$. For relatively high $q_{\min}$, the mode is identified as a reversed-shear Alfv\'en eigenmode (RSAE), characterized by its finite Alfv\'enic frequency and radial localization near the minimum of safety-factor profile. As $q_{\min}$ is further reduced, the dominant branch sequentially transitions through a low-frequency infernal-mode interval, then an energetic-particle-mode (EPM) regime, and finally another low-frequency infernal-mode interval. Increasing the EP beta fraction $\beta_h$ tends to destabilize the RSAE and EPM branches but to stabilize the infernal modes. Phase-space diagnostics further indicate that the destabilizing effects of EPs on the RSAE and EPM branches are mainly associated with trapped-particle drive, whereas the stabilizing effects of EPs on the infernal modes is dominated by passing particles. 

\end{abstract}

%
\vspace{2pc}
\noindent{\it Keywords}: RSAE, infernal mode, EPM, NIMROD, $q_{min}$
%

\submitto{\PPCF}
%
%
%
\pagestyle{fancy} 
\fancyhf{} 
\fancyhead[C]{\itshape {Alternation between infernal modes and Alfv\'en eigenmodes}} 
\renewcommand{\headrulewidth}{0.0pt}
\fancyfoot[C]{\thepage} 
\section{Introduction}

Advanced tokamak configurations commonly employ reversed or weak magnetic shear to improve confinement and enable high-performance steady-state operation, where the minimum safety factor, $q_{\min}$, is a key parameter governing both Alfv\'enic and pressure-driven instabilities. In particular, reversed-shear Alfv\'en eigenmodes (RSAEs), also referred to as Alfv\'en cascades, are discrete shear-Alfv\'en eigenmodes localized near the minimum of the safety-factor profile driven by the energetic particles (EPs). The RSAE frequency can sweep upward or downward as $q_{\min}$ evolves, making RSAEs useful indicators of current-profile evolution in reversed-shear plasmas \cite{berkTheoreticalInterpretationAlfven2001a,marchenkoAlfvenCascadesDownward2011}. This diagnostic capability, together with the relevance of RSAEs to energetic-particle transport, has motivated extensive experimental and numerical studies. Experimentally, RSAEs were first observed in reversed-shear discharges on JT-60U \cite{kimuraAlfvenEigenmodeEnergetic1998}, and were subsequently reported in JET \cite{sharapovAlfvenCascadesJET2006,kramerReversedShearAlfven2008}, DIII-D \cite{vanzeelandCouplingGlobalToroidal2007}, ASDEX Upgrade \cite{dagracaCharacterizationAlfvenEigenmodes2012}, Alcator C-Mod \cite{edlundExperimentalStudyReversed2010}, HL-2A \cite{chenDestabilizationReversedShear2014,chenCorelocalizedAlfvenicModes2016}, and EAST \cite{zhangExperimentalObservationReverse2018}. Numerical studies on RSAE have also been carried out using several simulation tools, including GTC \cite{dengLinearPropertiesRSAE2012,liuRegulationAlfvenEigenmodes2022}, NIMROD \cite{houNumericalStudyTransition2019}, MEGA \cite{wangInteractionEnergeticParticles2011a}, TAEFL \cite{spongSimulationAlfvenFrequency2013}, NOVA-K and TGLFEP \cite{zouValidationAlfvenEigenmode2019}.


{Infernal modes are low-frequency, pressure-driven MHD instabilities that arise in weak- or reversed-shear plasmas and are intrinsically global modes beyond the scope of conventional high-$n$ ballooning theory \cite{manickamIDEALMHDSTABILITY,charltonLinearNonlinearProperties1990a}. They are typically characterized by nearly zero real frequencies, localization in the weak/reversed-shear region, and non-monotonic dependences of the growth rate on $q_{\min}$ and toroidal mode number $n$.} Thus, in reversed-shear equilibria, variations of $q_{\min}$ can modify both the Alfv\'enic RSAE branch and the pressure-driven infernal branch. Previous studies have also connected infernal-type modes with energetic-ion dynamics. Circulating energetic ions were shown to drive infernal-fishbone-like energetic-particle modes \cite{kolesnichenkoInterchangeInfernalFishbone2006}, while saturated infernal modes can induce outward fast-ion transport \cite{marchenkoFastIonTransport2014}. Recent simulations have further examined kinetic-ion effects on ideal and resistive infernal modes \cite{satoKineticmagnetohydrodynamicHybridSimulation2024}, as well as resistive interchange, double-tearing, and infernal modes in negative-shear equilibria \cite{jardinIdealMHDLimited2022,jardinMHDStabilitySpherical2024,jardinResistiveInterchangeDouble2025}.

Although both RSAEs and infernal modes are strongly affected by weak or reversed magnetic shear and the structure of the safety-factor profile, they belong to different mode branches. Their competition and transition in reactor-relevant reversed-shear equilibria are less well-known. Variations of $q_{\min}$ can simultaneously modify the Alfv\'en continuum, the location of rational surfaces, and the pressure-driven stability boundary, allowing the dominant instability to switch between Alfv\'enic and low-frequency infernal branches. Moreover, EPs do not universally destabilize all branches: they can drive RSAE/EPM branches, but may provide a stabilizing kinetic response to infernal modes. A systematic study of these $q_{\min}$-organized branch transitions and branch-dependent EP effects is therefore essential for understanding stability boundaries in advanced tokamak plasmas.

In this work, we investigate the linear stability of toroidal mode number $n=4$ instabilities in a reversed-shear CFETR equilibrium \cite{zouValidationAlfvenEigenmode2019,zhuMHDAnalysisPhysics2022}. The Alfv\'en continuum is calculated using GTAW \cite{huNumericalStudyAlfven2014a}, while the mode frequencies, growth rates, and structures are obtained from NIMROD simulations \cite{sovinecNonlinearMagnetohydrodynamicsSimulation2004,kimHybridKineticMHDSimulations2004,kimImpactVelocitySpace2008}. By scanning $q_{\min}$, we identify a transition of the dominant mode from an RSAE branch to a higher-$q_{\min}$ infernal-mode interval, then to an EPM regime, and finally to a lower-$q_{\min}$ infernal-mode interval. Calculations show that the infernal modes persist in the absence of EP drive and are stabilized by the kinetic EP response. These results provide a unified overall picture of transitions between Alfv\'enic and infernal instabilities unique to the reversed-shear equilibria in advanced tokamak configurations.

The rest of paper is organized as follows. Section~\ref{simulation model} describes the hybrid kinetic-MHD model implemented in NIMROD. Section~\ref{simulation setup} presents the CFETR reversed-shear equilibrium, the initial EP distribution, and other numerical setups for the simulations. Section~\ref{results} discusses about the $q_{\min}$-induced transition among the RSAE, EPM, and infernal-mode regimes, as well as the branch-dependent role of EPs. Finally, Section~\ref{summary} summarizes the main results and discusses on their implications.

\section{Hybrid kinetic-MHD model}

\label{simulation model}
For the hybrid kinetic-MHD (HK-MHD) model implemented in the NIMROD code, the background plasma and energetic ions follow MHD equations and drift kinetic equations respectively \cite{sovinecNonlinearMagnetohydrodynamicsSimulation2004,kimHybridKineticMHDSimulations2004}. {In particular, the single-fluid ideal MHD equations are}
\begin{center}
	\begin{gather}
		\frac{\partial \rho}{\partial t}+\nabla \cdot(\rho \boldsymbol{V})=0 \\
		\rho\left(\frac{\partial \boldsymbol{V}}{\partial t}+\boldsymbol{V} \cdot \nabla \boldsymbol{V}\right)=\boldsymbol{J} \times \boldsymbol{B}-\nabla p_b-\nabla \cdot \boldsymbol{P}_h \\
		\frac{1}{\Gamma-1}\left(\frac{\partial p}{\partial t}+\boldsymbol{V} \cdot \nabla p\right)=-p \nabla \cdot \boldsymbol{V} \\
		\frac{\partial \boldsymbol{B}}{\partial t}=-\nabla \times \boldsymbol{E} \\
		\boldsymbol{J}=\frac{1}{\mu_0} \nabla \times \boldsymbol{B} \\
		\boldsymbol{E}+\boldsymbol{V} \times \boldsymbol{B}=0
	\end{gather}
\end{center}
where subscripts $b, h$ denote {the} bulk plasma and the hot particles, $\rho, \boldsymbol{V}$ are the mass density and the velocity of bulk plasma, neglecting the contribution of fast particles, $p$ is the total pressure of plasma including fast particles, $p_b$ is the pressure of bulk plasma, {$\boldsymbol{P}_h$} is the pressure tensor of fast ions, and {$\Gamma$} is {the} ratio of specific heats. {The rest of the symbol definitions are conventional.}

In the above HK-MHD model, it is assumed that the density of fast species is much lower than that of bulk plasmas but the fast species pressure is on the order of the bulk plasma pressure, i.e. ${n_h} \ll n_b$ and ${\beta_h} \sim \beta_b$, and $\beta \equiv 2 \mu_0 p / B^2$ is the ratio of thermal energy to magnetic energy \cite{chengKineticmagnetohydrodynamicModelLowfrequency1991a}. In this approximation, we neglect the contribution of energetic particles to the center of mass velocity. Assuming the center of the mass velocity of energetic ions to be zero, ${\boldsymbol{P}_h}$ in the momentum equation can be calculated from the velocity distribution function of energetic ions. The $\delta f$ PIC method is utilized to solve the drift kinetic equation for energetic ions {\cite{kimHybridKineticMHDSimulations2004}}.
\begin{equation}
	{ 
	\label{eq:drift1}
		 \dot{\boldsymbol{x}}=v_{\|} \hat{\boldsymbol{b}}+\frac{m}{e B^4}\left(v_{\|}^2+\frac{v_{\perp}^2}{2}\right)\left(\boldsymbol{B} \times \nabla \frac{B^2}{2}\right)+\frac{\boldsymbol{E} \times \boldsymbol{B}}{B^2}+\frac{\mu_0 m v_{\|}^2}{e B^2} \boldsymbol{J}_{\perp}} 
\end{equation}
\begin{equation}
	{ 
	\label{eq:drift2}
	m \dot{v}_{\|}=-\hat{\boldsymbol{b}} \cdot(\mu \nabla B-e \boldsymbol{E})}
\end{equation}
\noindent
{where $v_\perp$ ($v_\parallel$) is the velocity perpendicular (parallel) to the magnetic field, $\mu$ is the magnetic moment, $\hat{\boldsymbol{b}}=\boldsymbol{B}/B$ is the unit vector along the magnetic field, $m$ is the mass of the energetic particle, and $e$ is the electric charge.} {The individual terms in Eq.~\eqref{eq:drift1} correspond to the standard drift
	velocities in the drift–kinetic description. The first term,
	$v_{\|}\hat{\boldsymbol{b}}$, gives the parallel motion along the magnetic field.
	The second term,
	represents the combined curvature and $\nabla B$ drift. The third term corresponds to the 
	$\boldsymbol{E}\times\boldsymbol{B}$ drift. And the last term is the finite-pressure correction to the curvature and $\nabla B$ drifts, where $\boldsymbol{J}_{\perp} = 
	\boldsymbol{J}-\boldsymbol{J}\!\cdot\!\hat{\boldsymbol{b}}\,\hat{\boldsymbol{b}}$ \cite{kimImpactVelocitySpace2008}.} The energetic ion distribution function can be written as $f_h=f_{h0}+\delta f_{ h}$, where {$f_{h0}$} and {$\delta f_{ h}$} are the equilibrium and the perturbed components, and this gives {$\boldsymbol{P}_h=\boldsymbol{P}_{h0}+\delta \boldsymbol{P}_h$}, where {$\boldsymbol{P}_{{h0}}$} is assumed isotropic, and {$\delta \boldsymbol{P}_{{h}}$} is defined as 
\begin{equation}
	\label{eq:delta p}
	\delta {\boldsymbol{P}_{{h}}}=\left(\begin{array}{ccc}
		\delta p_{\perp} & 0 & 0 \\
		0 & \delta p_{\perp} & 0 \\
		0 & 0 & \delta p_{\|}
	\end{array}\right)
\end{equation}
{where $\delta p_{\perp} = \int \mu B \delta f_h d^3 v$ {($\delta p_\parallel = \int v_{\|}^2 \delta f_h d^3 v$)} is the {stress tensor component} due to hot ion motions perpendicular (parallel) to the magnetic field \cite{kimImpactVelocitySpace2008}}.

\section{Numerical setup}
\label{simulation setup}

The equilibrium used in this work is based on a CFETR steady-state scenario with a reversed-shear safety-factor profile \cite{zhuMHDAnalysisPhysics2022,zhouOptimizationsCFETRSteady2022}. The computational domain is bounded by the last closed flux surface (LCFS) and is discretized using a two-dimensional bicubic finite-element mesh aligned with the equilibrium magnetic flux surfaces, as shown in Figure~\ref{figure1}. The radial coordinate used in the following analysis is $\rho=\sqrt{\psi_N}$, where $\psi_N$ is the normalized poloidal flux. The main parameters of the reference equilibrium and EP distribution are summarized in Table~\ref{table:parameters}.

\begin{center}
    [Figure 1 about here.]\\
\end{center}

Because the reversed-shear profile contains two local minima, they are denoted by $q_{\min,1}$ and $q_{\min,2}$, with $q_{\min,1}<q_{\min,2}$. Unless otherwise specified, $q_{\min}$ in the following refers to $q_{\min,1}$. This scan is used to follow the transition of the dominant $n=4$ instability among the RSAE, EPM, and infernal-mode regimes.

Energetic ions are initialized using an isotropic slowing-down distribution \cite{kimImpactVelocitySpace2008},
\begin{equation}
\label{eq:slowing_down}
f_{h0}
=
\frac{P_0\exp(P_{{\phi}}/\psi_n)}
{{E}^{3/2}+{E}_c^{3/2}},
\end{equation}
where $P_0$ is a normalization constant, $P_{{\phi}}=g\rho_\parallel-\psi_p$ is the canonical toroidal momentum, $g=RB_\phi$, $\rho_\parallel=mv_\parallel/(eB)$, $\psi_p$ is the poloidal flux, and $\psi_n=c\psi_0$ controls the radial localization of the EP profile. Here, ${E}$ is the fast ion energy and ${E}_c$ is the critical slowing-down energy,
\begin{equation}
\label{eq:critical_energy}
{E}_c
=
\left(\frac{3}{4}\right)^{2/3}
\left(\frac{\pi m_i}{m_e}\right)^{1/3}T_e ,
\end{equation}
where $m_i$ and $m_e$ are the ion and electron masses, and $T_e$ is the electron temperature.

In the $\beta_h$ scan, the EP beta replaces the corresponding fraction of the bulk-plasma beta, while the total equilibrium beta profile is kept fixed. This treatment allows the change in mode growth rate to be attributed to the kinetic EP response rather than to a modification of the total equilibrium beta drive. Unless otherwise specified, the reference EP beta fraction is $\beta_h=0.43$.

\begin{table}[htbp]
    \centering
    \caption{Main simulation parameters for the reference CFETR equilibrium and EP distribution.}
    \label{table:parameters}
    \begin{tabular}{|l|c|}
        \hline
        \textbf{Parameter} & \textbf{Value} \\
        \hline
        Major radius, $R$ & 7.2 m \\
        Minor radius, $a$ & 2.2 m \\
        Toroidal magnetic field at magnetic axis, $B_0$ & 6.5 T \\
        Electron temperature at magnetic axis, $T_{e0}$ & 45.6 keV \\
        Reference EP beta fraction, $\beta_h$ & 0.43 \\
        \hline
    \end{tabular}
\end{table}

\section{Simulation results and analyses}
\label{results}

We first examine the stability and mode structure of the reference CFETR reversed-shear equilibrium. This reference case provides a baseline for identifying the Alfv\'enic branch before the minimum safety factor $q_{\min}$ is varied. The dominant toroidal mode number considered in this work is $n=4$.

\subsection{Reference RSAE identification}
\label{rsae_case}

At the reference equilibrium, with $q_{\min}=1.27$ and $\beta_h=0.43$, the dominant $n=4$ instability is first analyzed in terms of its mode structure and interaction with the Alfv\'en continuum. Figure~\ref{fig:reference_rsae}(a) shows the two-dimensional mode structure in terms of the flow velocity component $V_{\Psi}$ normal to the flux surface. The corresponding poloidal Fourier spectrum (PFS) is shown in Figure~\ref{fig:reference_rsae}(b), with the dominant $m=4-6$ components highlighted, while other harmonics are plotted in gray. The mode is radially localized in the region between the two local minima of the safety factor profile, $q_{\min,1}$ and $q_{\min,2}$, where the magnetic shear is weak.

\begin{center}
    [Figure 2 about here.]\\
\end{center}

The Alfv\'en continuum, calculated using GTAW, is shown in Figure~\ref{fig:reference_rsae}(c) as red curves, with the NIMROD-calculated mode frequency indicated by a blue dashed marker. The mode frequency lies inside the TAE gap and is well separated from the continuum branches, indicating that the mode is a discrete Alfv\'enic eigenmode. The mode does not exhibit a clear two-harmonic structure typical of conventional TAE modes, but is instead dominated by a single poloidal harmonic ($m=6$). Radially, the mode profile is peaked and localized around the center of the weak magnetic shear region between the inner negative shear ($\rho \lesssim 0.22$) and the outer positive shear ($\rho \gtrsim 0.53$) regions. These features are reminiscent of an RSAE.

\subsection{Mode transitions as $q_{\min}$ is varied}
\label{mode_transitions}

We next examine how the dominant $n=4$ instability evolves as the minimum safety factor $q_{\min}$ (specifically $q_{\min,1}$) is varied. Figure~\ref{fig:n4_scan} shows the real frequency and growth rate as functions of $q_{\min}$ at $\beta_h = 0.43$. The scan indicates that the dominant instability is strongly influenced by $q_{\min}$ and can be divided into three main regimes: low-frequency infernal mode for $q_{\min}\le 1.04$ and $1.17 < q_{\min} \le 1.25$, EPM at intermediate $q_{\min}$, and RSAE for $q_{\min}> 1.25$. The classification here is primarily based on the real frequency: infernal modes have $f < 10~\mathrm{kHz}$, EPM in the $10$--$20~\mathrm{kHz}$ range (well below the TAE gap), and RSAE at much higher frequencies. This frequency-based classification provides a clear distinctions among the mode branches, and the identification of the low-frequency branches as infernal modes is further verified below using calculations without EPs.
\begin{center}
    [Figure 3 about here.]\\
\end{center}
Representative two-dimensional mode structures and corresponding PFS for selected $q_{\min}$ intervals are shown in Figure~\ref{fig:n4_structures_combined}. The infernal-mode and EPM structures are visualized using the perturbed pressure, whereas the RSAE structure is represented using $V_\psi$ in Figure~\ref{fig:reference_rsae}(a). Although both EPM and infernal modes reside in the reversed-shear region, their spatial characteristics differ: infernal modes are more localized near the $q_{\min,1}$ region, whereas EPMs exhibit a broader radial extent with several comparable poloidal components. The growth rates of both infernal modes and EPM show non-monotonic behavior as functions of $q_{\min}$. These observations provide a classification of the $n=4$ instability across the variation range of $q_{\min}$. Definitive identification of the low-frequency branches as infernal modes will be performed in the next subsection via calculations without EP effects.
\begin{center}
    [Figure 4 about here.]\\
\end{center}

\subsection{Identification of the low-frequency branches as infernal modes}
\label{infernal_id}

To further identify the low-frequency branches observed in the $q_{\min}$ scan, we perform calculations in the absence of EPs, while the equilibrium profiles are kept unchanged. In the absence of EPs, the ideal-MHD operator is self-adjoint, and the unstable modes exhibit nearly zero real frequency ($\omega_r \simeq 0$). Therefore, the frequency evolution versus $q_{\min}$ is not shown here, and the mode identification is based on the growth rate behavior and mode structures.

\begin{center}
    [Figure 5 about here.]\\
\end{center}

Figure~\ref{fig:no_ep_scan} shows the growth rate of the dominant $n=4$ instability as a function of $q_{\min}$ without EP effects, together with the corresponding EP case for comparison. The low-frequency branches remain unstable in the absence of EPs, and their growth rates are larger than those obtained with EPs. By contrast, the intermediate-$q_{\min}$ EPM branch and the higher-$q_{\min}$ RSAE branch are strongly stabilized. This result indicates that the low-frequency branches are not EP-driven modes, but rather MHD instabilities of the reversed-shear equilibrium.

\begin{center}
    [Figure 6 about here.]\\
\end{center}

To further illustrate their spatial characteristics, Figure~\ref{fig:infernal_contours_noep} presents the two-dimensional contour plots of the perturbed pressure for four representative cases, $q_{\min}=0.98$, $1.02$, $1.20$, and $1.24$, all calculated without EPs. In all these cases, the mode structures are localized in the weak- or reversed-shear region, consistent with the expected localization of infernal modes in advanced tokamak equilibria. From the contour plots in Figure~\ref{fig:infernal_contours_noep} and the growth rate scan in Figure~\ref{fig:no_ep_scan}, it can be seen that the mode growth rate reaches a maximum at $q_{\min,1}=0.98$ with $m=4$ and at $q_{\min,2}=1.18$ with $m=5$ respectively (Fig.~\ref{fig:infernal_contours_noep}a). Similarly, for $q_{\min,1}=1.02$ and $q_{\min,2}=1.22$, the $m=4$ and $m=5$
components are localized near the inner and outer minima respectively {(Fig.~\ref{fig:infernal_contours_noep}b)}, all consistent with $m\sim nq_{\min}$ at both locations.

\begin{center}
    [Figure 7 about here.]\\
\end{center}

Another important feature of infernal modes is the non-monotonic, oscillatory dependence of the growth rate on toroidal mode number $n$ \cite{manickamIDEALMHDSTABILITY,charltonLinearNonlinearProperties1990a}. To verify this property, we scan $n$ for two representative cases, $q_{\min}=1.02$ and $q_{\min}=1.24$. The growth rates shown in Figure~\ref{fig:n_scan_infernal} indeed demonstrate the oscillatory behavior. These features are all consistent with the established properties of infernal modes in low- or reversed-shear tokamak plasmas \cite{manickamIDEALMHDSTABILITY,charltonLinearNonlinearProperties1990a}. The low-frequency branches observed in the $q_{\min}$ scan are identified as infernal modes.

\subsection{Energetic-particle stabilization of infernal modes}
Having identified the two low-frequency branches as infernal modes, we next examine how EPs affect their stability. For this purpose, $\beta_h$ is varied for two representative infernal-mode cases, $q_{\min}=1.02$ and $q_{\min}=1.24$.

\begin{center}
    [Figure 8 about here.]\\
\end{center}

\noindent{}{Figure~\ref{fig:betah_scan_infernal} shows the growth rates of the two representative infernal modes as functions of $\beta_h$. In both cases, the growth rate decreases monotonically as $\beta_h$ increases. Since the equilibrium $\beta$ profile is unchanged in the $\beta_h$ scan, the reduction represents a net stabilizing contribution from the kinetic EP response.}

\begin{center}
    [Figure 9 about here.]\\
\end{center}

To identify the EP population associated with each mode
branch, Figure~\ref{fig:phase_space_ep} shows
$|\delta f_4|/|\delta f_4|_{\max}$, the normalized magnitude of the
$n=4$ component of the EP distribution perturbation, evaluated within
the radial interval where each mode is localized and projected onto
the $(v_\parallel,v_\perp)$ plane. {Four representative cases are considered: the infernal modes at $q_{\min}=0.98$ and $1.24$, the EPM at $q_{\min}=1.10$, and the RSAE at $q_{\min}=1.27$.} The red dotted lines provide the median of the local trapped-passing boundaries evaluated within the selected radial intervals between the inner and outer $q_{min}$ locations. In Figure~\ref{fig:phase_space_ep}(a) and (c), the strongest responses of the two infernal-mode cases lie mainly on the passing-particle side of the boundary, whereas the EPM and RSAE responses in
Figure~\ref{fig:phase_space_ep}(b) and (d) are concentrated
predominantly on the trapped-particle side. This phase-space
distinction motivates the examination of passing-particle resonances
for the infernal branches and trapped-particle resonances for the EPM
and RSAE branches.

For passing particles, the transit resonance condition
is~\cite{heidbrinkBasicPhysicsAlfven2008}

\begin{equation}
  \omega \simeq n\omega_\phi-m\omega_\theta ,
  \label{eq:passing_resonance}
\end{equation}
where $\omega_\phi$ and $\omega_\theta$ are the toroidal and poloidal
transit frequencies. For trapped particles, the bounce--precession
resonance condition is~
\cite{porcelliSolutionDriftKinetic1994,
fredricksonBouncePrecessionFishbones2003}
\begin{equation}
  \omega \simeq n\omega_d+l\omega_b ,
  \label{eq:trapped_resonance}
\end{equation}
where $\omega_d$ is the toroidal precession frequency,
$\omega_b$ is the bounce frequency, and $l$ is the bounce harmonic number. {Accounting for the finite resonance width associated with the linear growth rate~\cite{briguglioAnalysisNonlinearBehavior2014,zoncaNonlinearDynamicsPhase2015,wangStructureWaveParticle2016}, resonant EPs are selected according to

\begin{equation}
  {|\Omega_{\mathrm{res}}-\omega|}
  \leq \gamma,
  \label{eq:resonance_detuning}
\end{equation}}

\noindent where $\Omega_{\mathrm{res}}=4\omega_\phi-m\omega_\theta$ for passing
particles and $\Omega_{\mathrm{res}}=4\omega_d+l\omega_b$ for trapped particles. {The corresponding resonant region in velocity space is denoted as $\mathcal{R}(v_\parallel,v_\perp)$.}

\begin{center}
	[Figure 10 about here.]\\
\end{center}

{The contribution of the resonant EPs to the $n=4$ distribution perturbation is evaluated as}

\begin{equation}
	|\delta f_{4,\mathrm{res}}(v_\parallel,v_\perp)|
	=
	\left|
	\sum_{p\in\mathcal{R}(v_\parallel,v_\perp)}
	\delta f_p e^{i4\phi_p}
	\right| ,
	\label{eq:resonant_n4_projection}
\end{equation}

\noindent where $\delta f_p$ is the contribution of marker $p$ to the EP
distribution perturbation, and $\phi_p$ is the toroidal location of marker $p$. Figure~\ref{fig:coherent_contribution_ep} presents the resulting normalized $n=4$ projections of the resonant EP perturbations. The dominant structures overlap
the high-amplitude regions in Figure~\ref{fig:phase_space_ep}. The corresponding resonances are also confirmed in $(P_\phi,E)$ space.
Figure~\ref{fig:pphi_energy_resonance} shows the normalized $n=4$
perturbed EP distributions in $(P_\phi,E)$ space for {various specific} pitch angles characterized by $\Lambda=\mu B_0/E$, together with the corresponding resonance lines. For the $q_{\min}=0.98$ and $1.24$ infernal modes, the perturbed EP distributions are concentrated around the $m=4$ and $m=5$ passing-particle resonance regions $\omega=4\omega_\phi-4\omega_\theta$ and $\omega=4\omega_\phi-5\omega_\theta$, respectively, whereas those for the $q_{\min}=1.10$ EPM and $q_{\min}=1.27$ RSAE are concentrated around the $l=-1$ and $l=2$ trapped-particle resonance {regions $\omega=4\omega_d-\omega_b$ and $\omega=4\omega_d+2\omega_b$, respectively}.

\begin{center} 
	{[Figure 11 about here.]}\\ 
\end{center}

\section{Conclusions and discussion}
\label{summary}

In this work, we have investigated the linear stability of the toroidal mode number $n=4$ instability in a reversed-shear CFETR equilibrium in the presence of energetic particles. The main objective is to clarify how the dominant mode branch changes with the minimum safety factor $q_{\min}$ and how energetic particles modify the stability of the selected branch. The reference case with $q_{\min}=1.27$ and $\beta_h=0.43$ is identified as an RSAE-type gap mode, due to its finite Alfv\'enic frequency, dominant $m=6$ poloidal harmonic, and localization in the weak-shear region between the two {neighbouring surfaces with strong negative and positive shears}. For relatively high $q_{\min}$, $1.25<q_{\min}<1.28$, the dominant mode remains in the RSAE regime. With decreasing $q_{\min}$, the system enters an EPM regime in the interval $1.04<q_{\min}\le1.17$. Low-frequency infernal-mode regimes are found for $q_{\min}\le1.04$ and in the interval $1.17<q_{\min}\le1.25$. These infernal modes are characterized by nearly zero real frequency, localization in the weak/reversed-shear region, and non-monotonic dependences of the growth rate on $q_{\min}$ and toroidal mode number. These low-frequency modes persist and exhibit larger growth rates in absence of EPs, indicating that they are not driven by energetic particles.

EPs are found to play a branch-dependent role. For the RSAE and EPM branches, EPs provide the destabilizing drive. In contrast, for the infernal-mode branches, increasing $\beta_h$ reduces the growth rate. Since the equilibrium $\beta$ profile is kept unchanged during the $\beta_h$ scan, this reduction is attributed to the kinetic EP response rather than to a modification of the equilibrium $\beta$ drive. Phase-space diagnostics suggest that different particle populations are involved in the RSAE and infernal-mode branches. The normalized $|\delta f_4|$ distributions show that the strongest responses of the RSAE and EPM branches are concentrated predominantly in the trapped-particle region, whereas those of the infernal modes lie mainly in the passing-particle region. The velocity-space distributions and $n=4$ projections of
the resonant EPs further support the identification of
passing-particle transit resonances for the infernal branches and
trapped-particle bounce--precession resonances for the EPM and RSAE
branches. {And the corresponding perturbed EP distributions in $(P_\phi,E)$ space are concentrated around these resonance lines.}

These findings suggest that stability assessment in reactor-relevant reversed-shear plasmas should consider not only energetic-particle-driven Alfv\'enic modes, but also the possible kinetic stabilization of pressure-driven infernal modes. Future work should extend the present linear analysis to nonlinear simulations, direct wave--particle power-transfer diagnostics, and multi-$n$ mode interactions.

\section*{Data availability statement}
All data that support the findings of this study are included within the article (and any supplementary files).

\section*{Acknowledgements}

We are grateful for the supports from the NIMROD team. This work is supported by the National MCF Energy R\&D Program of China Grant No.~2019YFE03050004, and the U.S. Department of Energy Grant No.~DE-FG02-86ER53218. The computing work in this paper is supported by the Public Service Platform of High Performance Computing by Network and Computing Center of HUST{, and this research used resources of the National Energy Research Scientific Computing Center, a DOE Office of Science User Facility supported by the Office of Science of the U.S. Department of Energy under Contract No. DE-AC02-05CH11231 using NERSC award FES-ERCAP0027638.}

\makeatletter
\setcounter{figure}{0}
\renewcommand{\thefigure}{\arabic{figure}}
\makeatother
\clearpage

\begin{figure}[H]
	\centering
	\subfloat[\label{flux.fig}]{\includegraphics[width=0.45\textwidth,height=7.5cm]{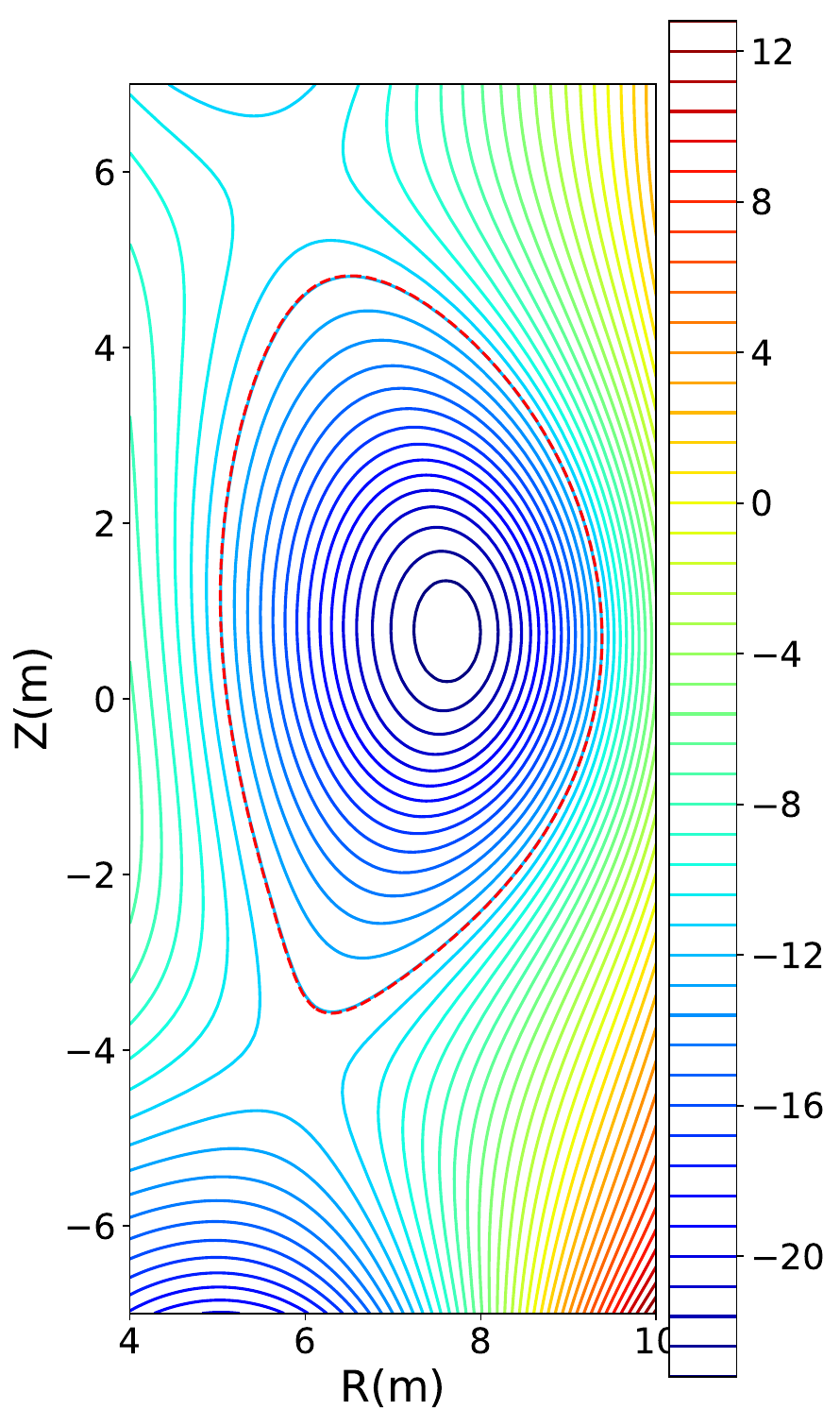}} 
	\vspace{0.2cm}
	\subfloat[\label{grid.fig}]{\includegraphics[width=0.4\textwidth,height=8cm]{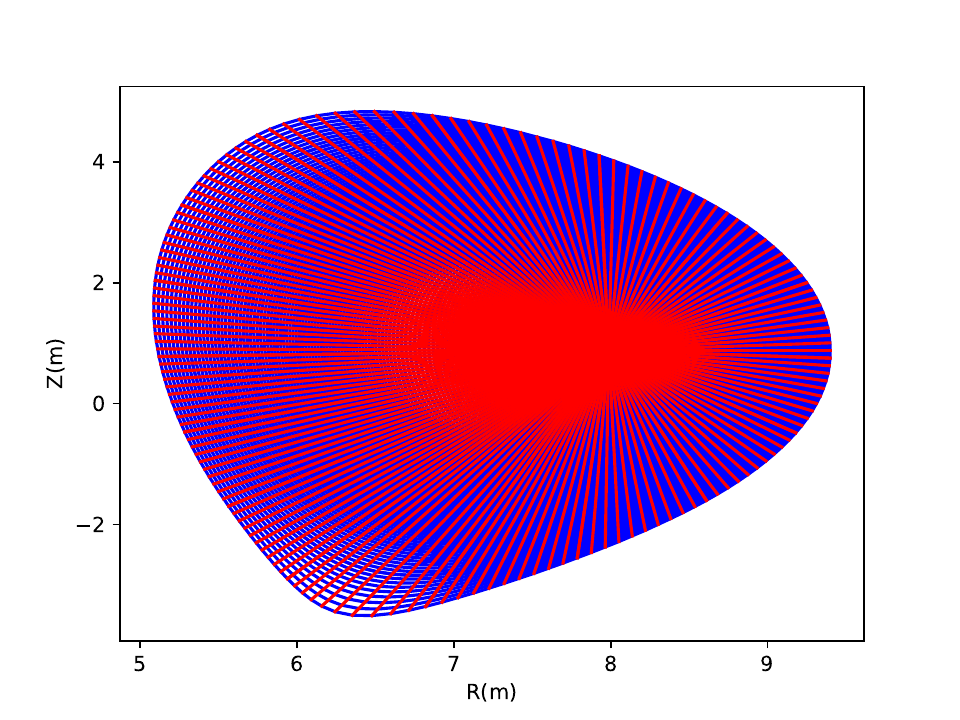}}\\
	\subfloat[\label{fig.safety}]{\includegraphics[width=0.45\textwidth,height=6.5cm]{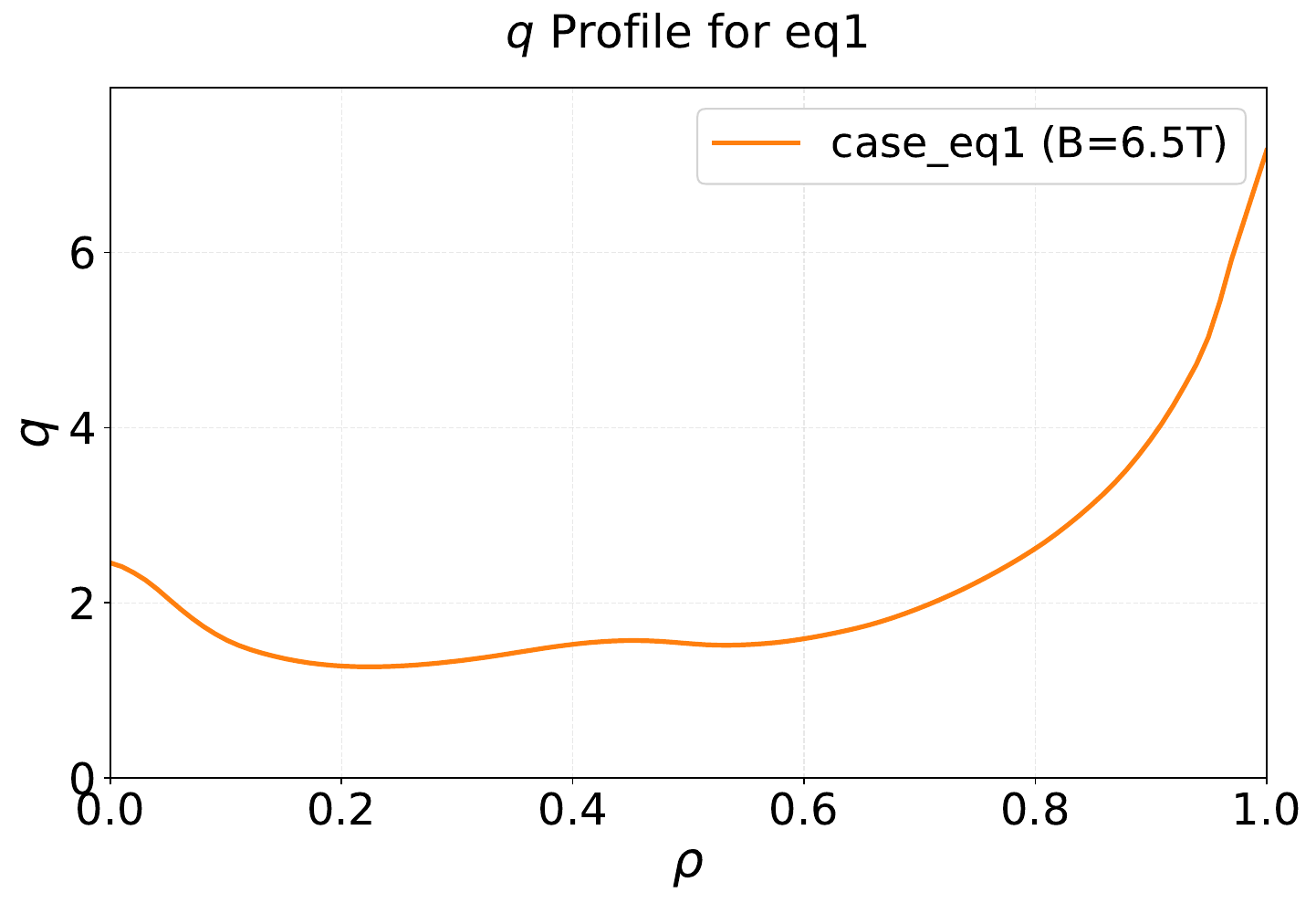}} 
	\vspace{1cm}
	\subfloat[\label{fig.pres}]{\includegraphics[width=0.45\textwidth,height=6.5cm]{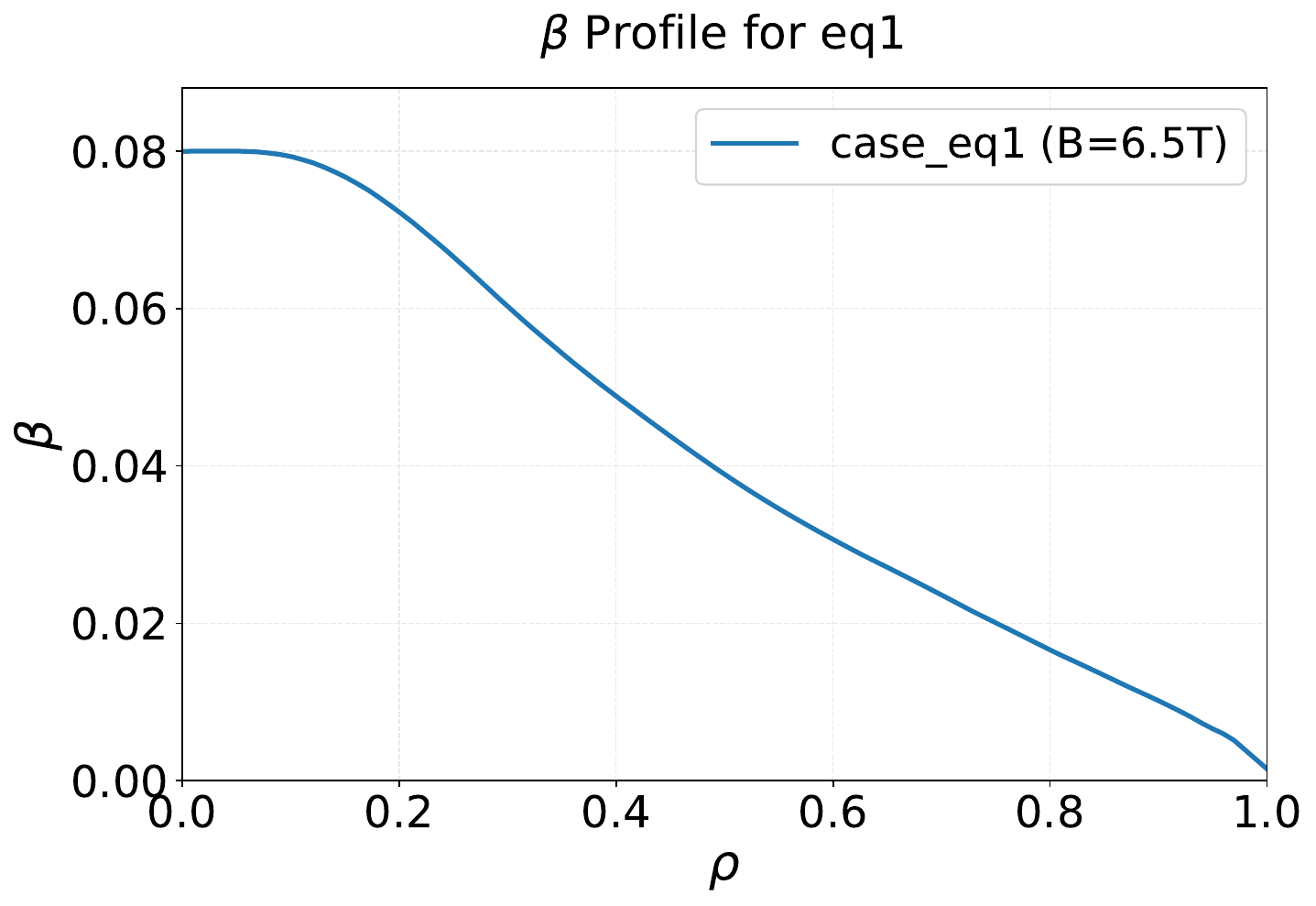}}
	\caption{(a) Contour plot of equilibrium poloidal flux in (R, Z) coordinate. The red curve represents the last closed flux surface (LCFS). (b) The mesh grid of flux coordinates used in the calculation. The blue lines represent the constant poloidal fluxes, and the red lines the poloidal angles. {(c) and (d) are the 1D radial profiles of safety factor and equilibrium $\beta$. The radial coordinate $\rho$ represents the square root of the normalized poloidal flux, and the minimum value of the \(q\) profile is specified as \(q_{\text{min}} = 1.27\). }}
	\label{figure1}
\end{figure}
\clearpage

\begin{figure}[H]
    \centering
    
    \subfloat[2D $V_\psi$ contour from NIMROD \label{fig:con_n4}]{
        \includegraphics[width=0.5\textwidth,height=7.5cm]{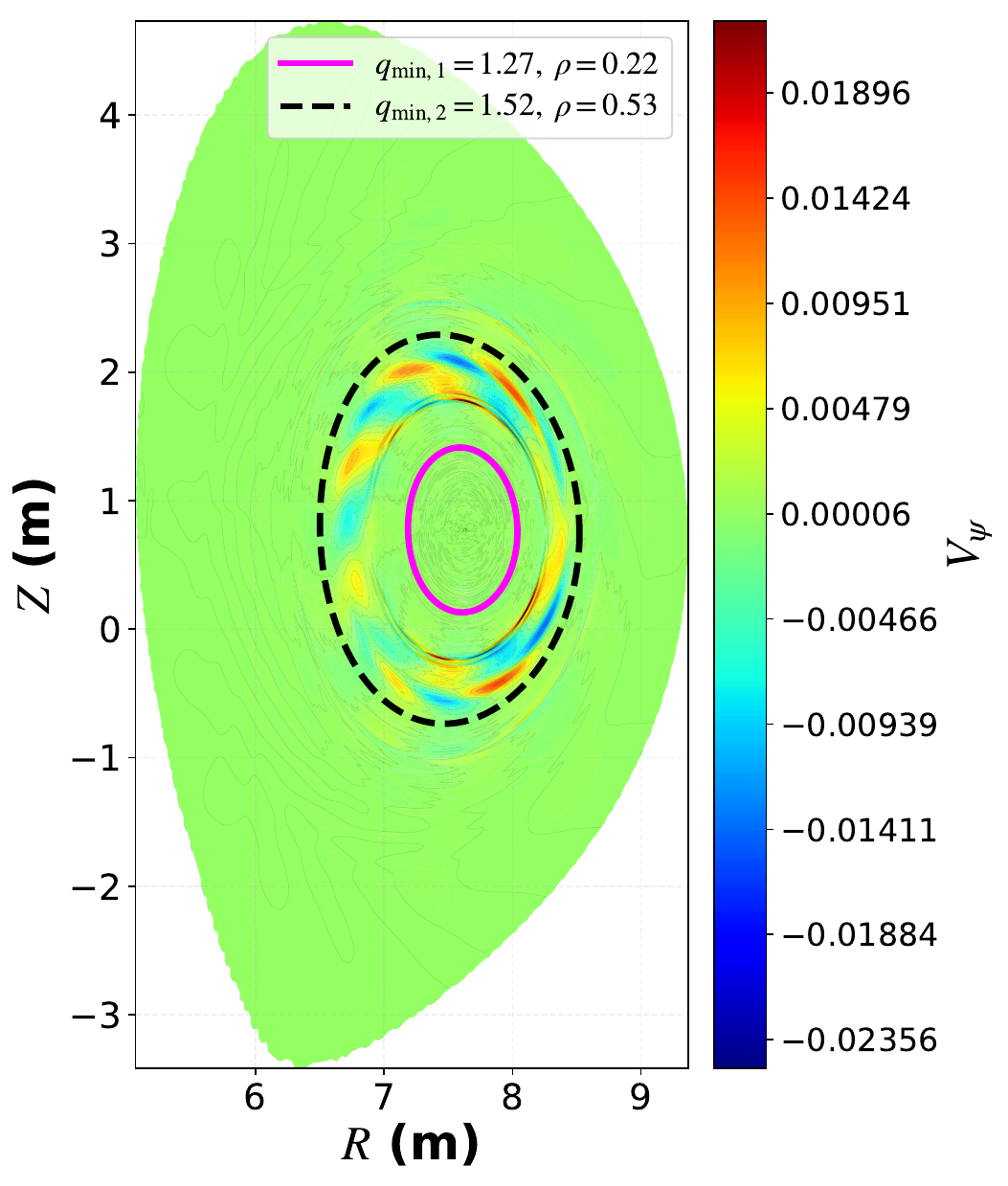}
    }
    \vfill
    
    \subfloat[PFS from NIMROD \label{fig:pfs_n4}]{
        \includegraphics[width=0.48\textwidth,height=5cm]{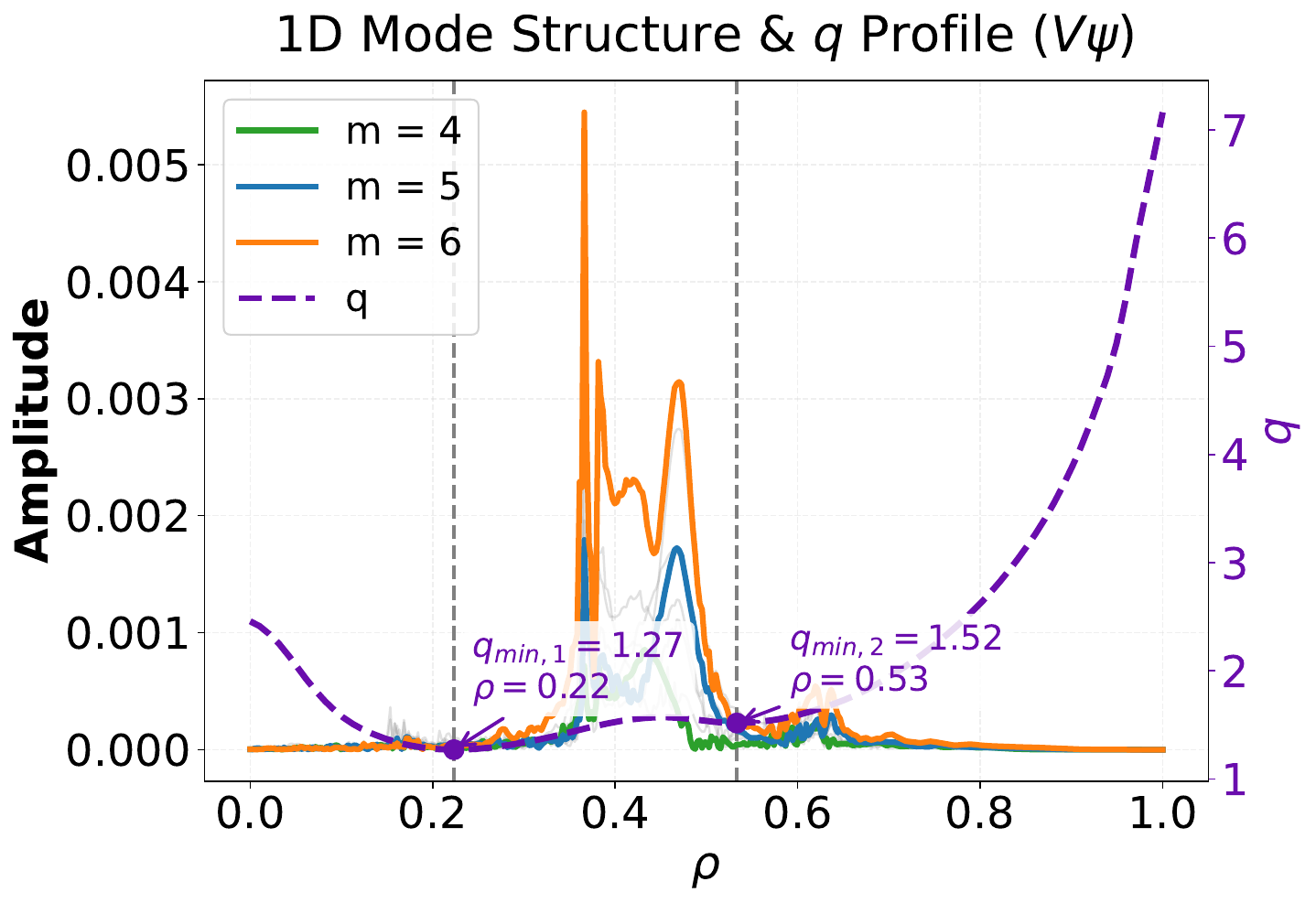}
    }\hfill
    \subfloat[Alfv\'en continuum from GTAW with NIMROD mode frequency \label{fig:alfven_continuum}]{
        \includegraphics[width=0.48\textwidth,height=5cm]{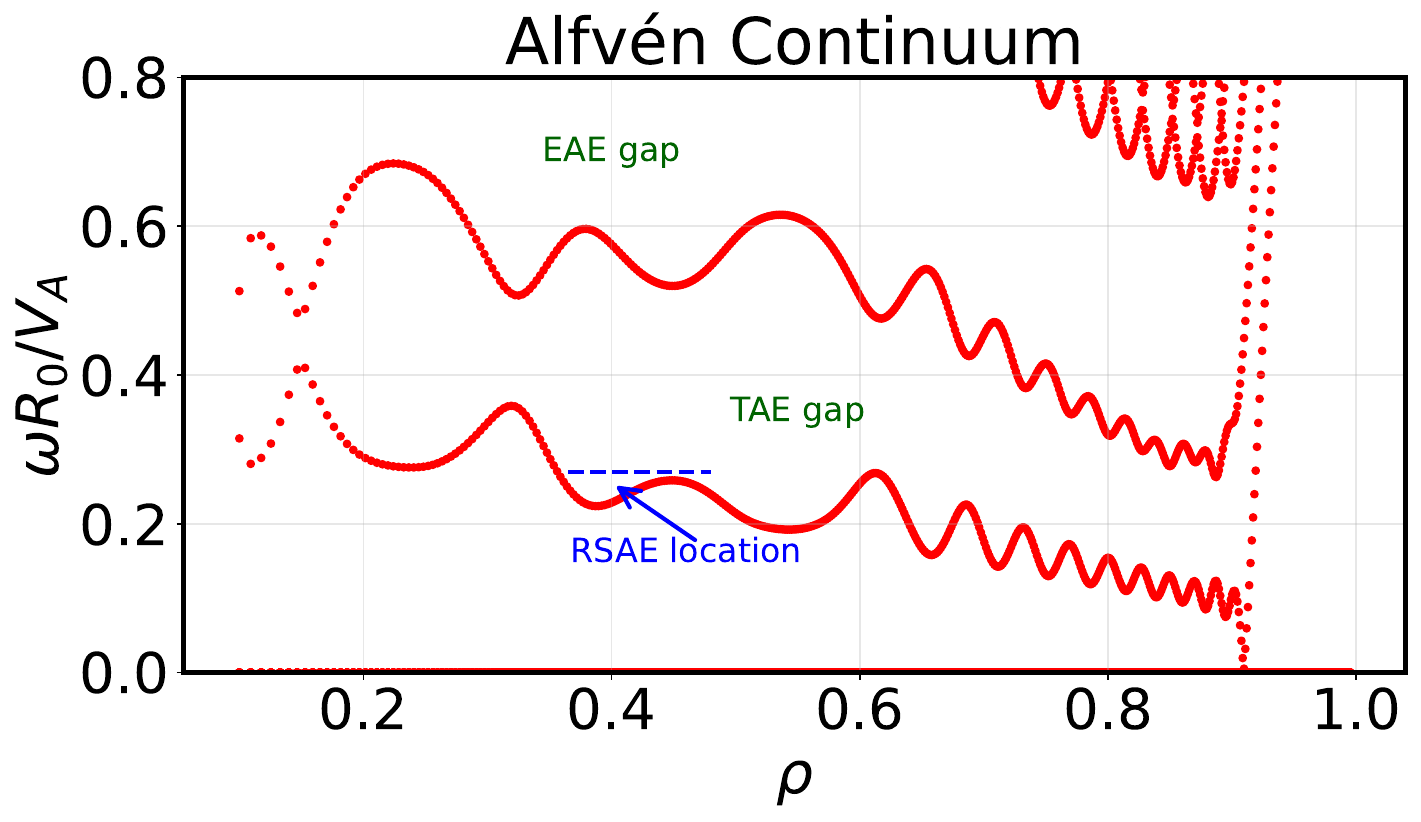}
    }
    \caption{
        Reference RSAE for the $n=4$ mode at $\beta_h=0.43$. 
        (a) Two-dimensional $V_\psi$ contour in poloidal plane. (b) Profiles of the dominant poloidal harmonics and the (c) corresponding Alfv\'en continuum along with the RSAE frequency.
    }
    \label{fig:reference_rsae}
\end{figure}
\clearpage

\begin{figure}[htbp]
    \centering
    \subfloat[Frequency vs $q_{\min}$]{%
        \includegraphics[width=0.85\textwidth]{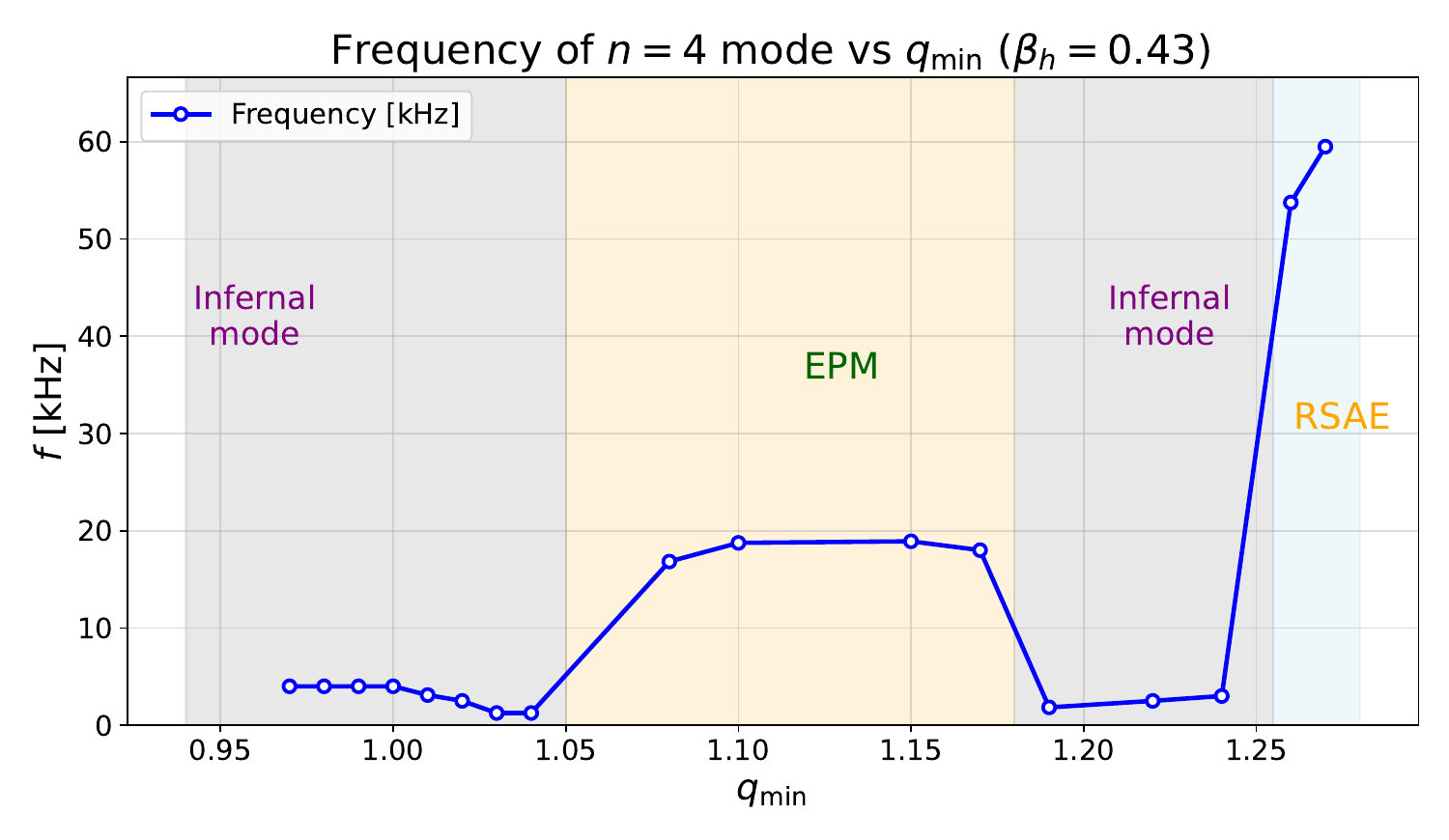}%
        \label{fig:n4_freq}
    }\\
    \subfloat[Growth rate vs $q_{\min}$]{%
        \includegraphics[width=0.85\textwidth]{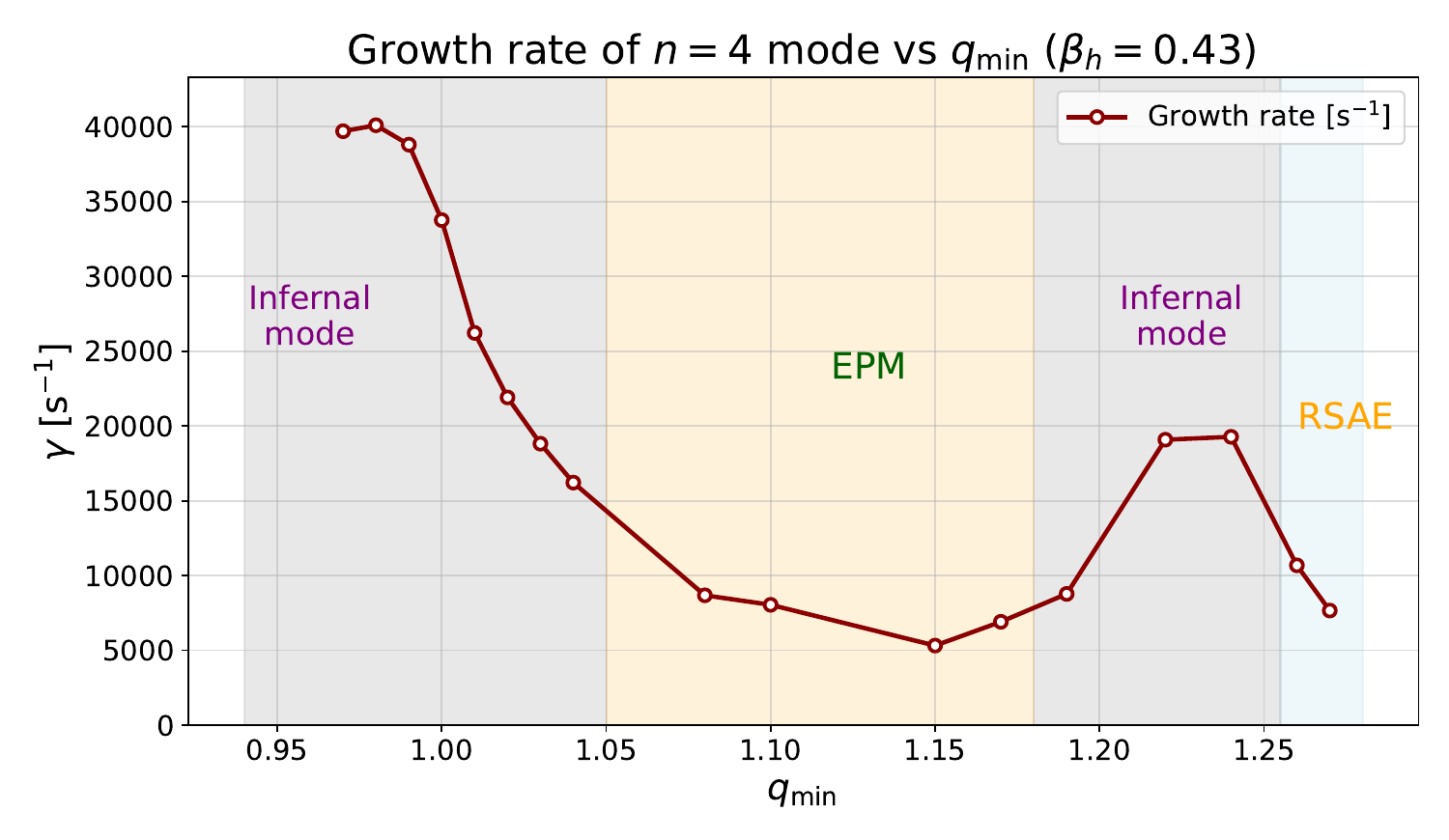}%
        \label{fig:n4_gamma}
    }
    \caption{
    Dependence of (a) the real frequency (kHz) and (b) the growth rate (s$^{-1}$) of the dominant $n=4$ instability on $q_{\min}$ at $\beta_h = 0.43$. Shaded regions indicate the infernal-mode, EPM, and RSAE regimes.}
    \label{fig:n4_scan}
\end{figure}
\clearpage

\begin{figure}[htbp]
    \centering

    \subfloat[Infernal mode: $q_{\min}=1.04$]{
        \includegraphics[width=0.40\textwidth,height=5cm]{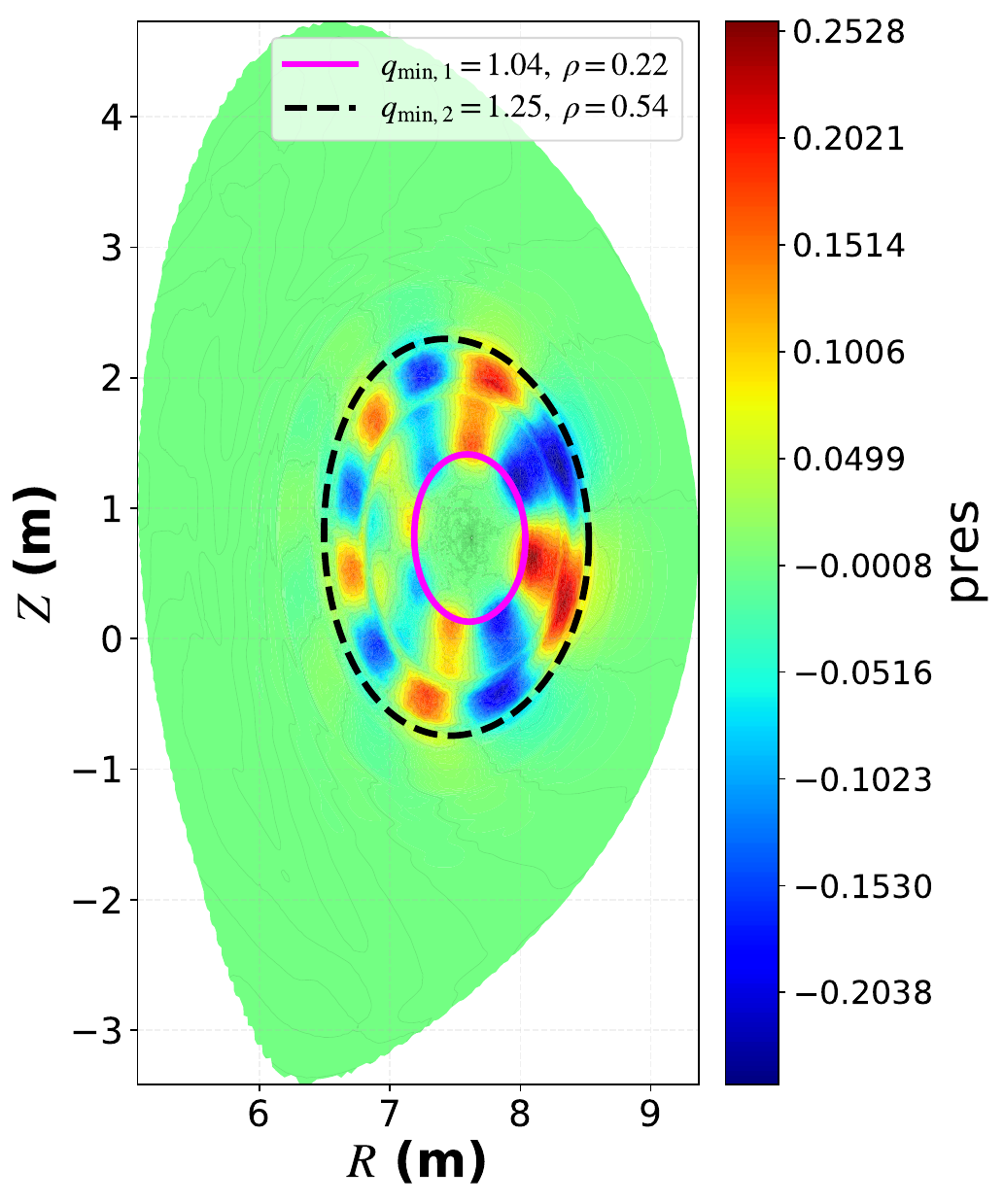}
        \label{fig:infernal_low}
    }
    \subfloat[Infernal mode: PFS, $q_{\min}=1.04$]{
        \includegraphics[width=0.45\textwidth]{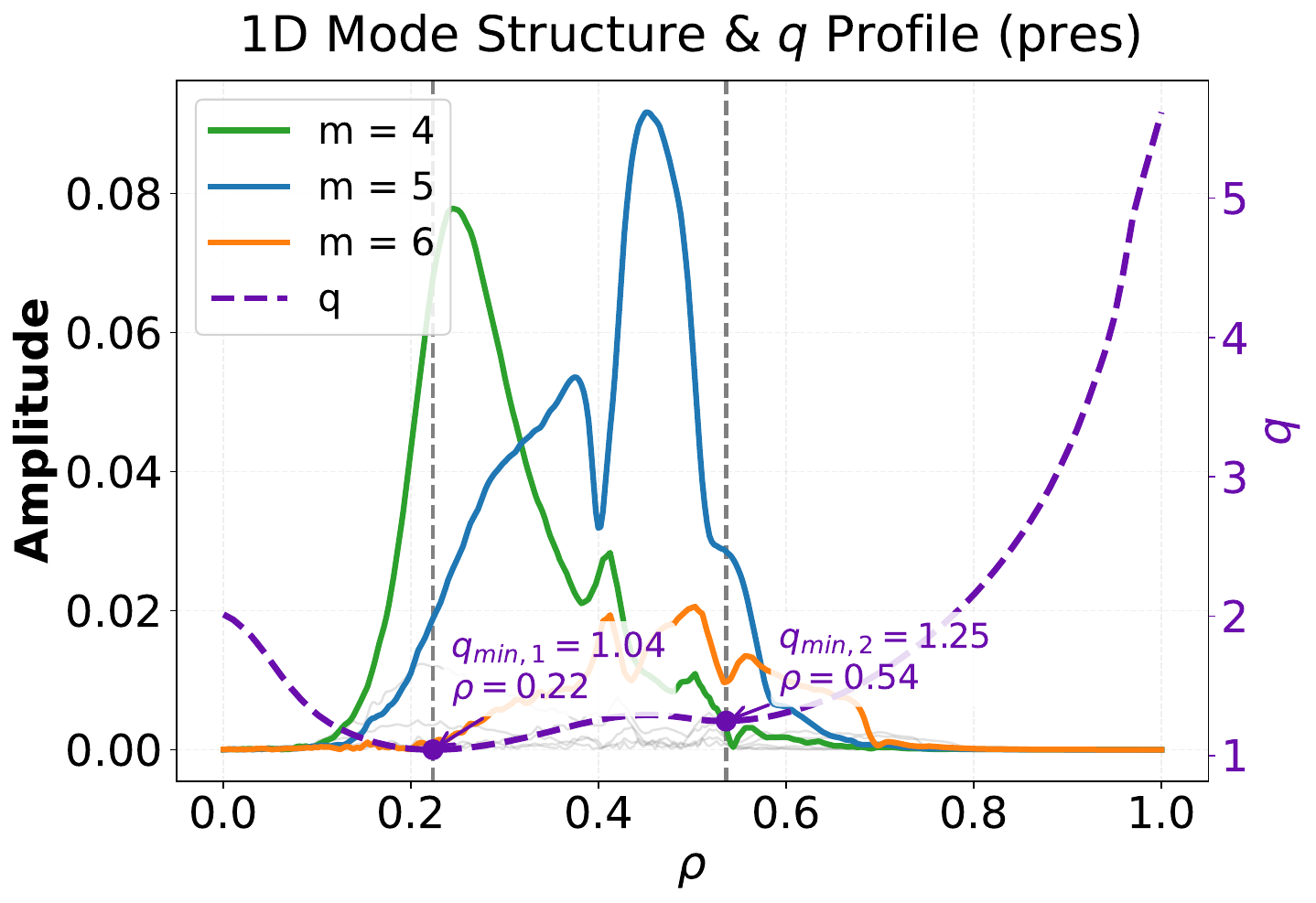}
        \label{fig:pfs_infernal_low}
    }\\[0.3cm]

    \subfloat[EPM: $q_{\min}=1.10$]{
        \includegraphics[width=0.40\textwidth,height=5cm]{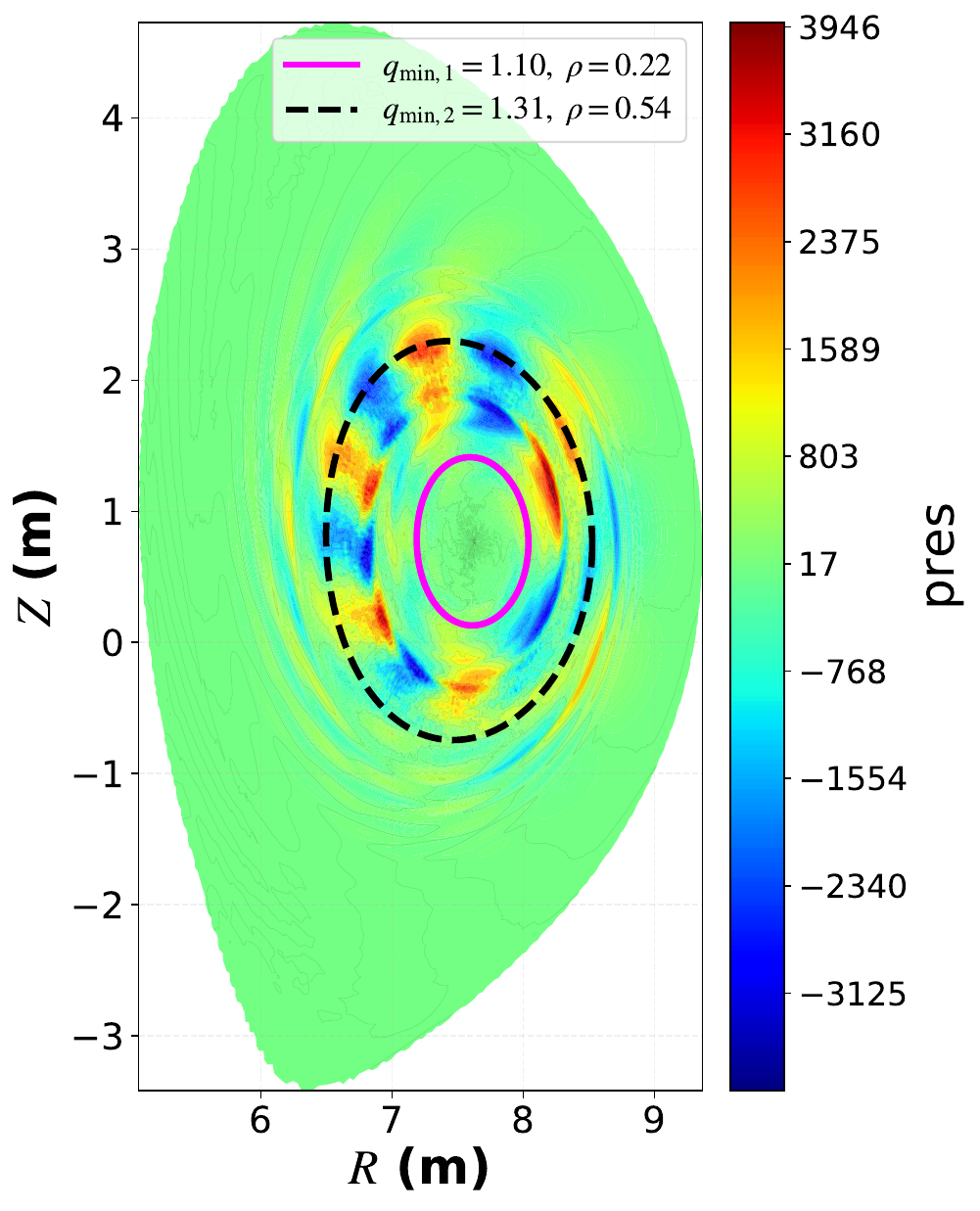}
        \label{fig:epm}
    }
    \subfloat[EPM: PFS, $q_{\min}=1.10$]{
        \includegraphics[width=0.45\textwidth]{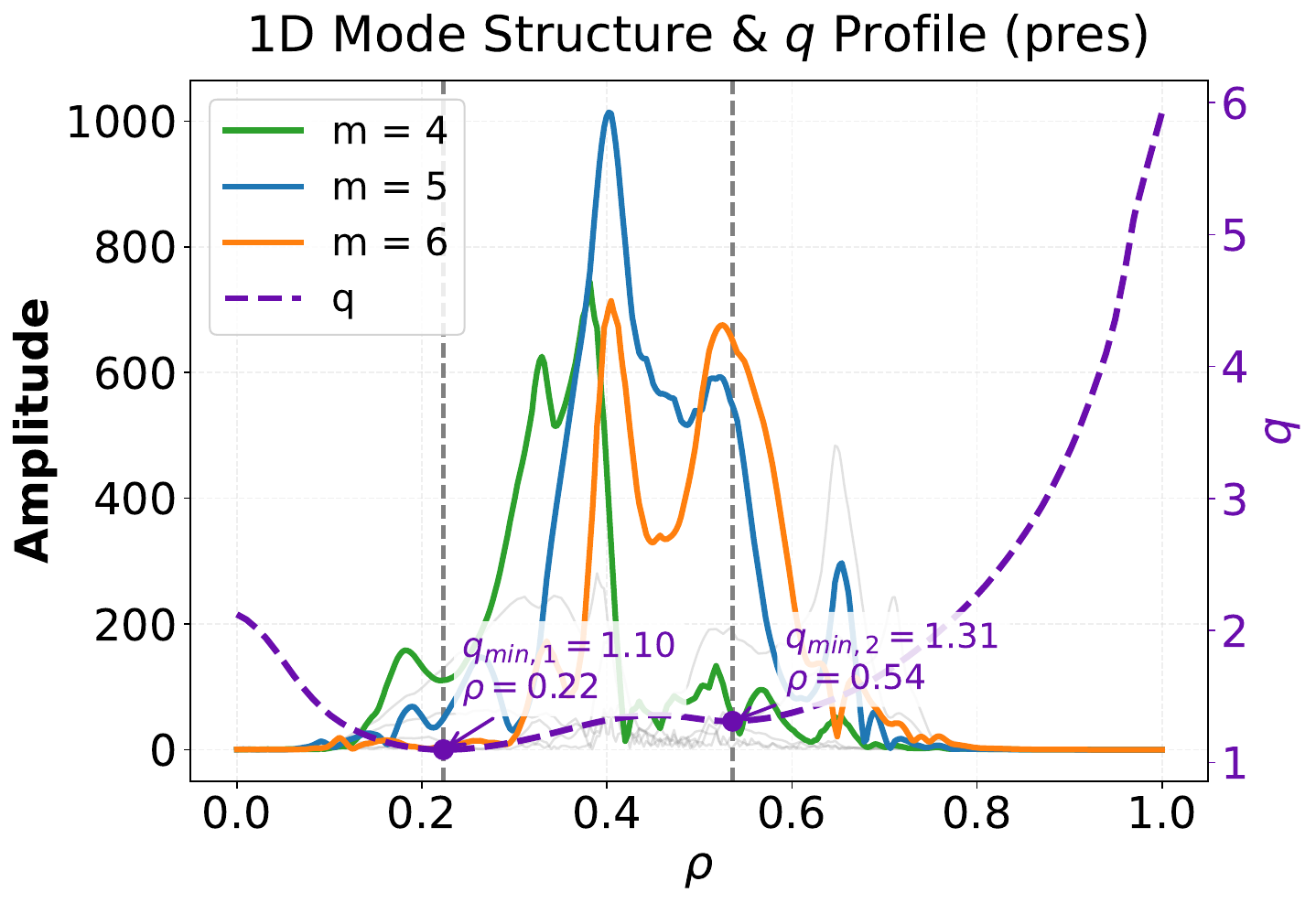}
        \label{fig:pfs_epm}
    }\\[0.3cm]

    \subfloat[Infernal mode: $q_{\min}=1.23$]{
        \includegraphics[width=0.40\textwidth,height=5cm]{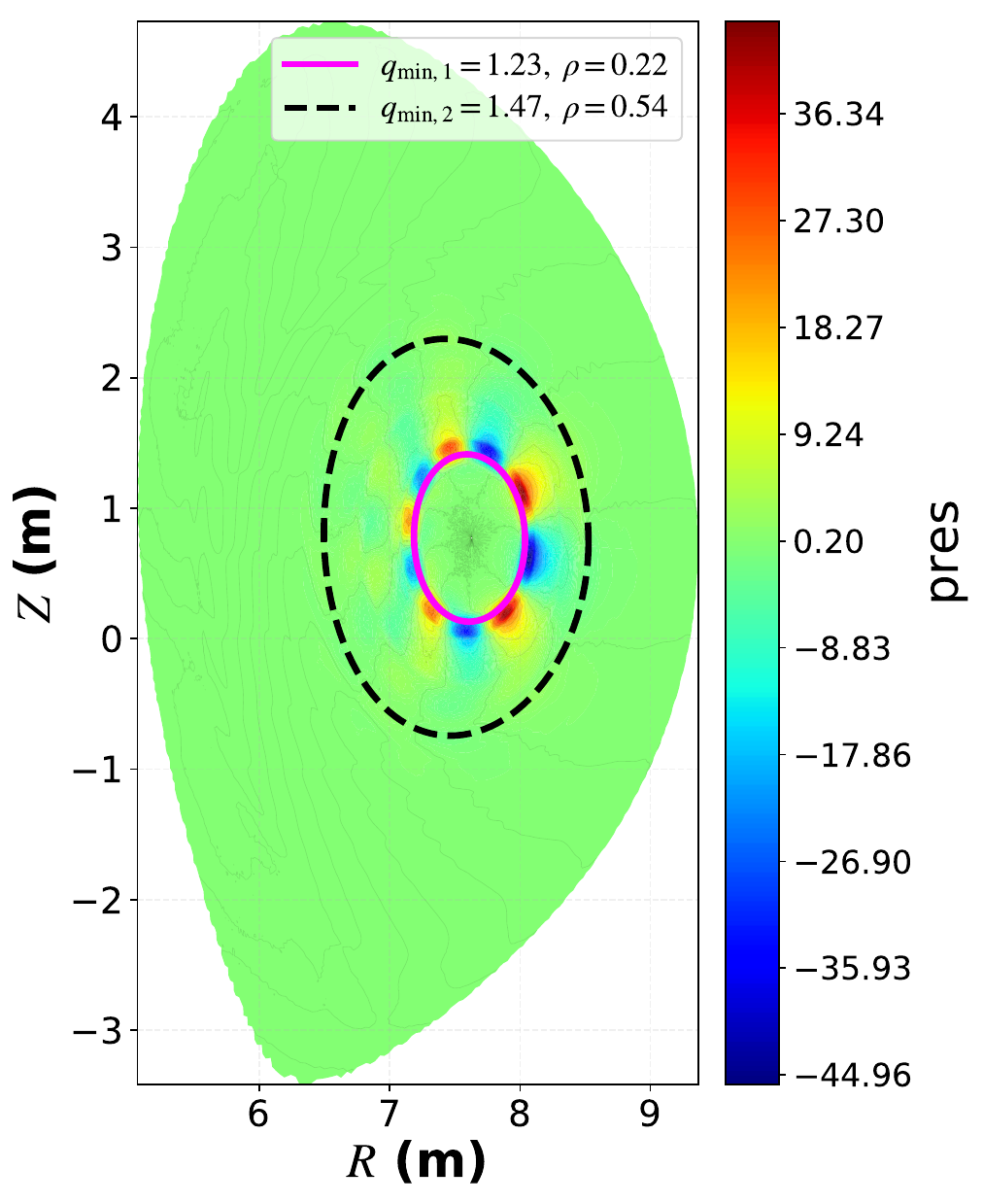}
        \label{fig:infernal_high}
    }
    \subfloat[Infernal mode: PFS, $q_{\min}=1.23$]{
        \includegraphics[width=0.45\textwidth]{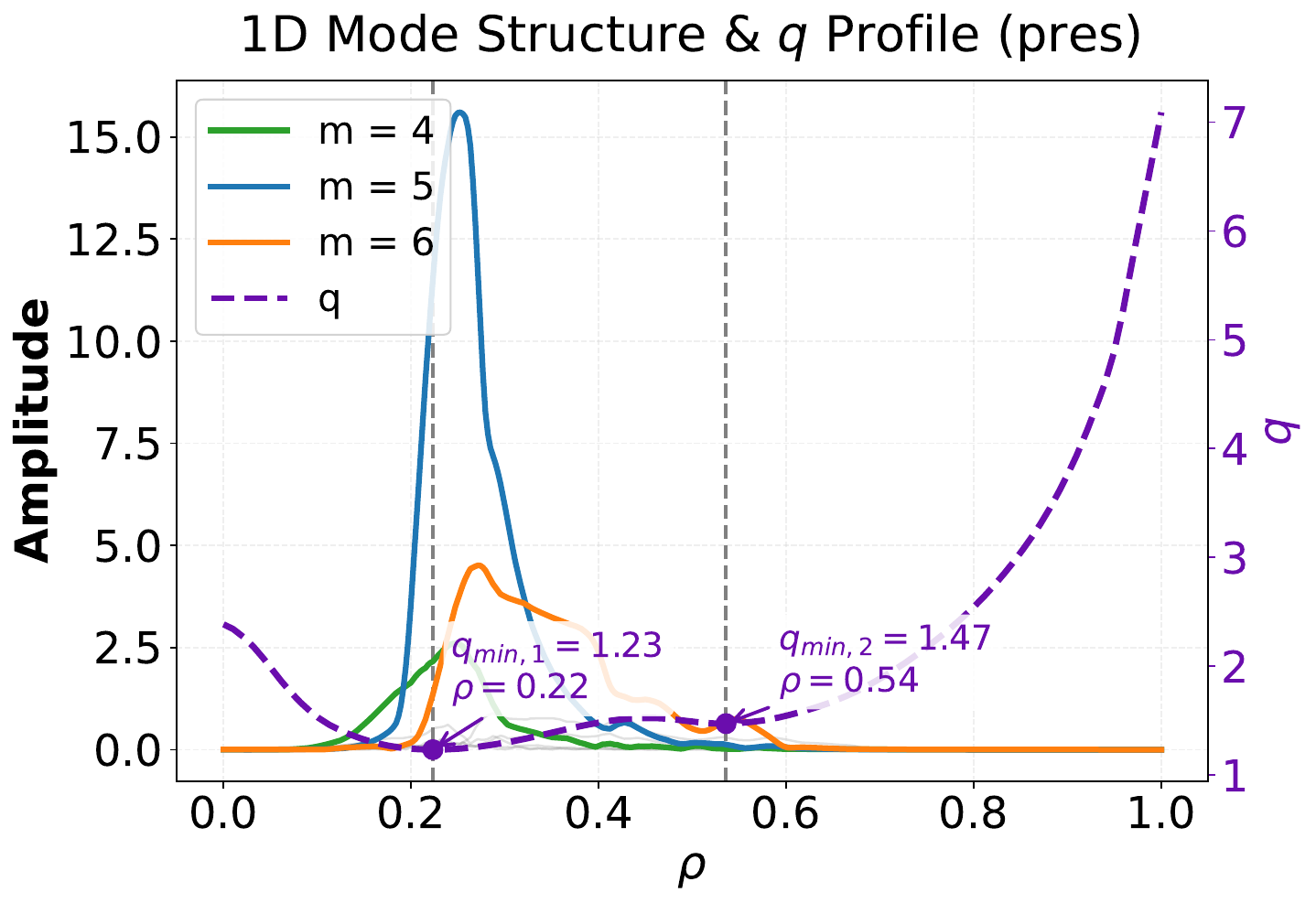}
        \label{fig:pfs_infernal_high}
    }

    \caption{
        Representative mode structures and corresponding PFS of the $n=4$ instability for the selected $q_{\min}$ intervals. Left column: perturbed pressure; right column: PFS. Top to bottom: low-$q_{\min}$ infernal mode ($q_{\min}=1.04$), intermediate-$q_{\min}$ EPM ($q_{\min}=1.10$), higher-$q_{\min}$ infernal mode ($q_{\min}=1.23$).
    }
    \label{fig:n4_structures_combined}
\end{figure}
\clearpage

\begin{figure}[htbp]
    \centering
    {%
        \includegraphics[width=0.85\textwidth]{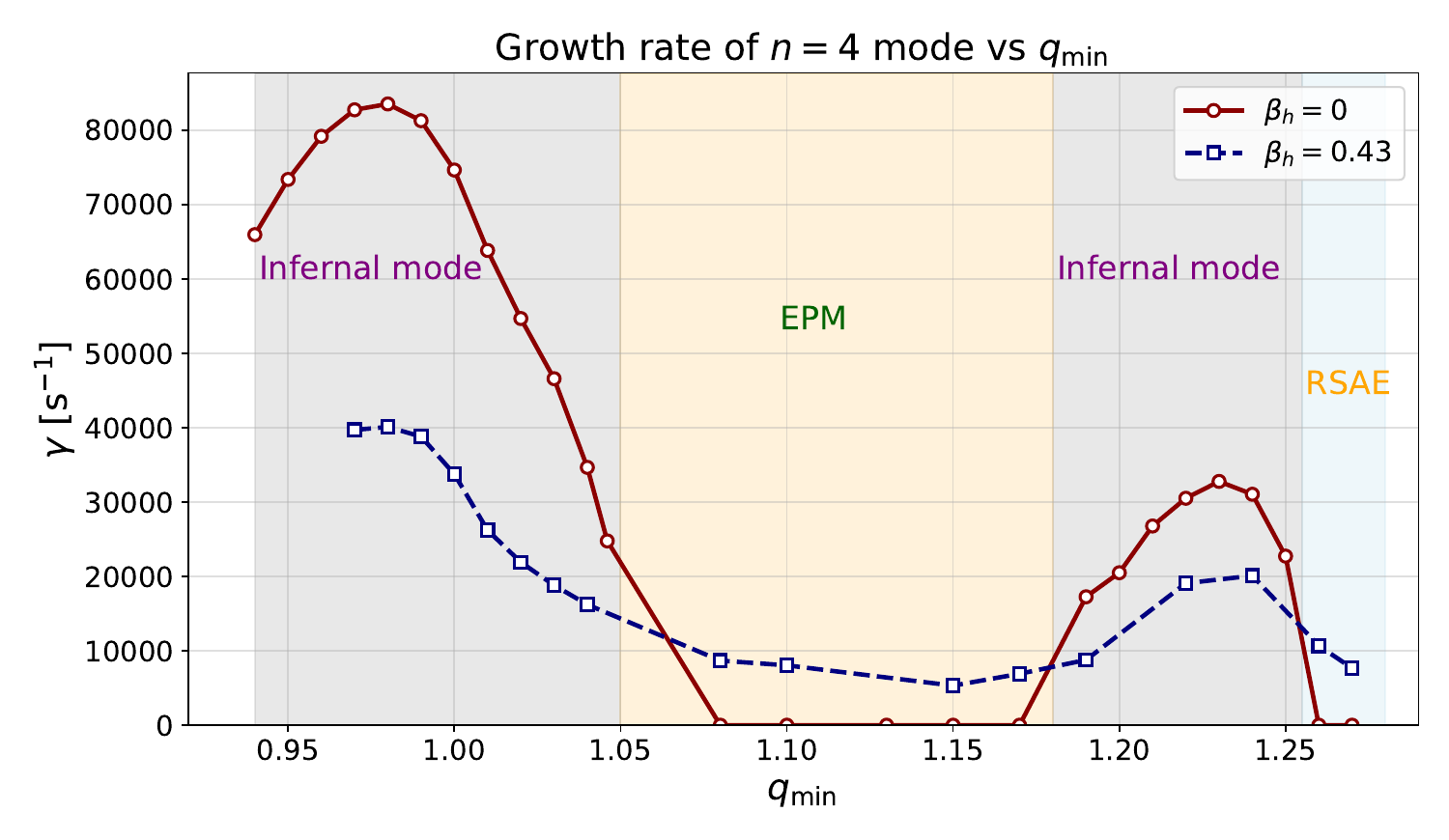}%
        \label{fig:qmin_scan_no_ep}
    }
    \caption{
    Growth rates of the $n=4$ instability versus $q_{\min}$ in absence ($\beta_h=0$) and in presence ($\beta_h=0.43$) of EPs.}
    \label{fig:no_ep_scan}
\end{figure}
\clearpage

\clearpage
\begin{figure}[htbp]
    \centering

    \subfloat[$q_{\min}=0.98$]{
        \includegraphics[width=0.45\textwidth]{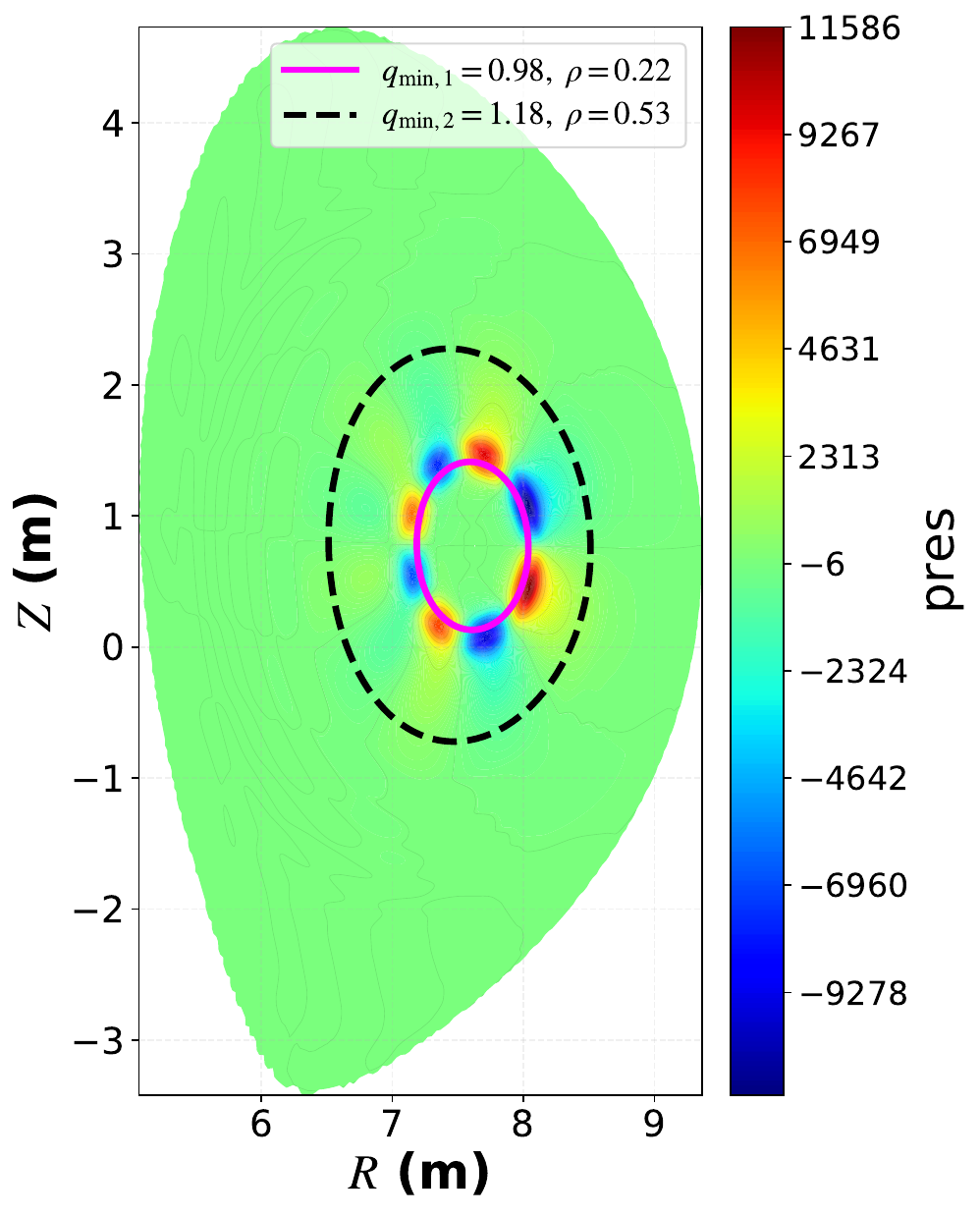}
        \label{fig:contour_q098_noep}
    }
    \subfloat[$q_{\min}=1.02$]{
        \includegraphics[width=0.45\textwidth]{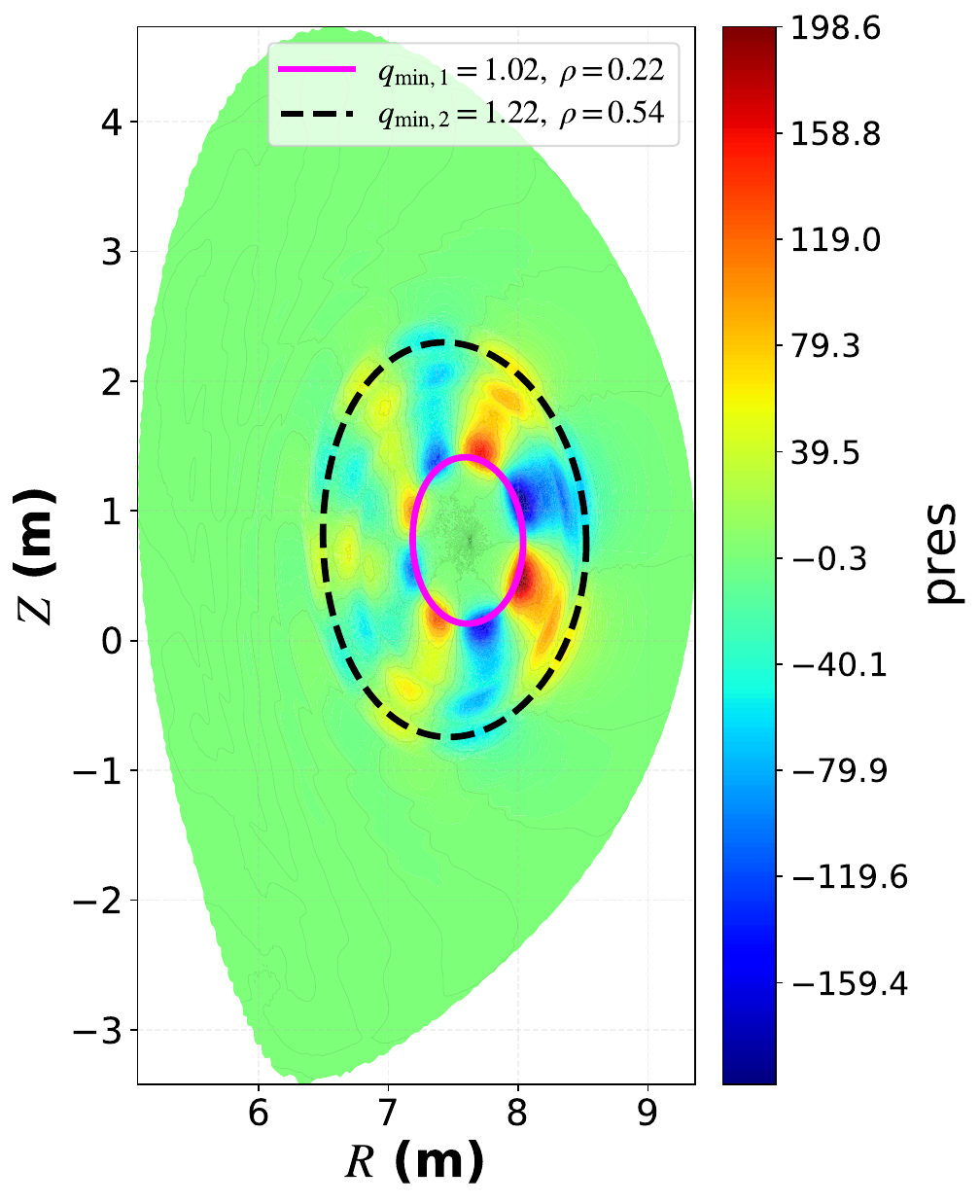}
        \label{fig:contour_q102_noep}
    }\\[0.3cm]

    \subfloat[$q_{\min}=1.20$]{
        \includegraphics[width=0.45\textwidth]{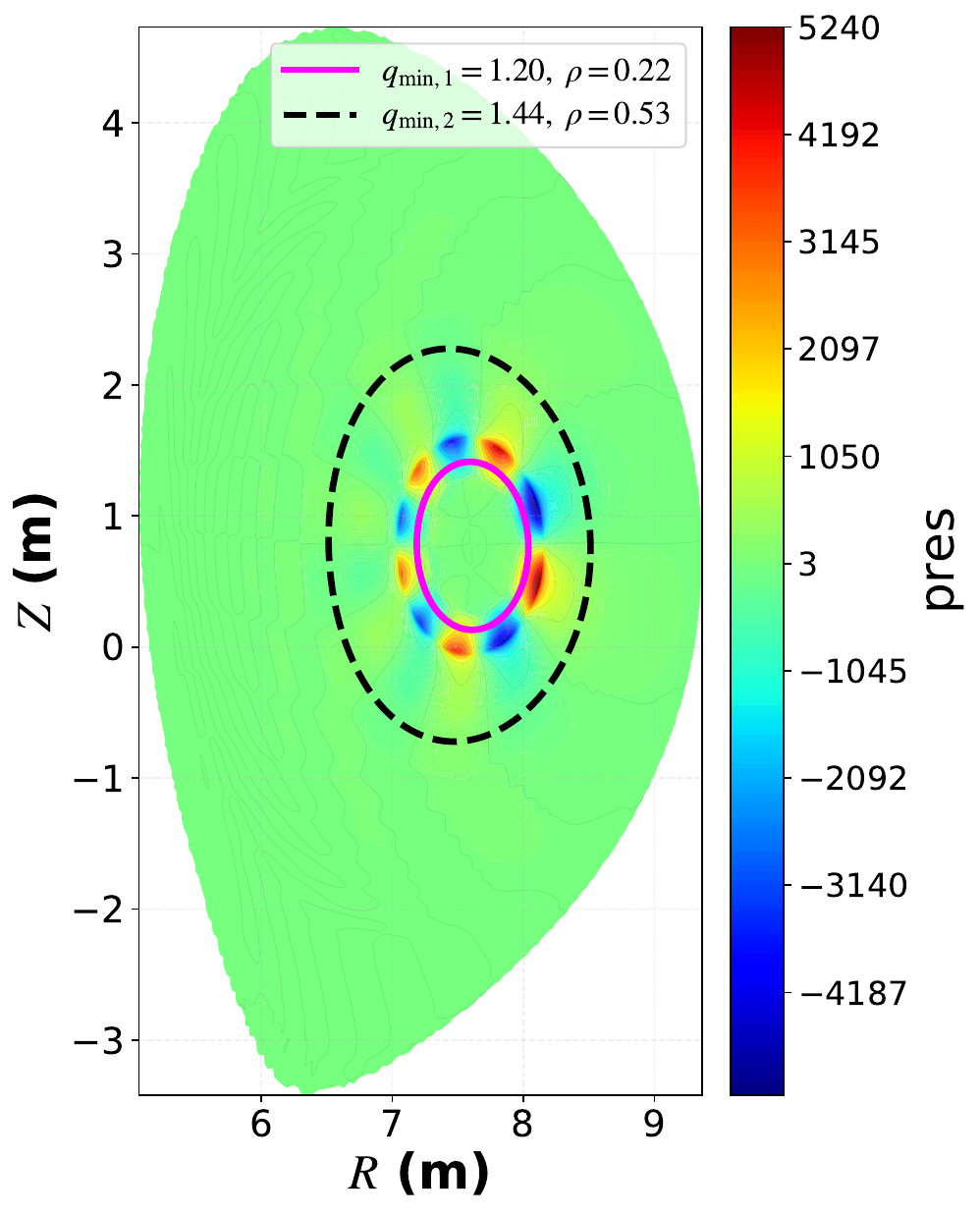}
        \label{fig:contour_q120_noep}
    }
    \subfloat[$q_{\min}=1.24$]{
        \includegraphics[width=0.45\textwidth]{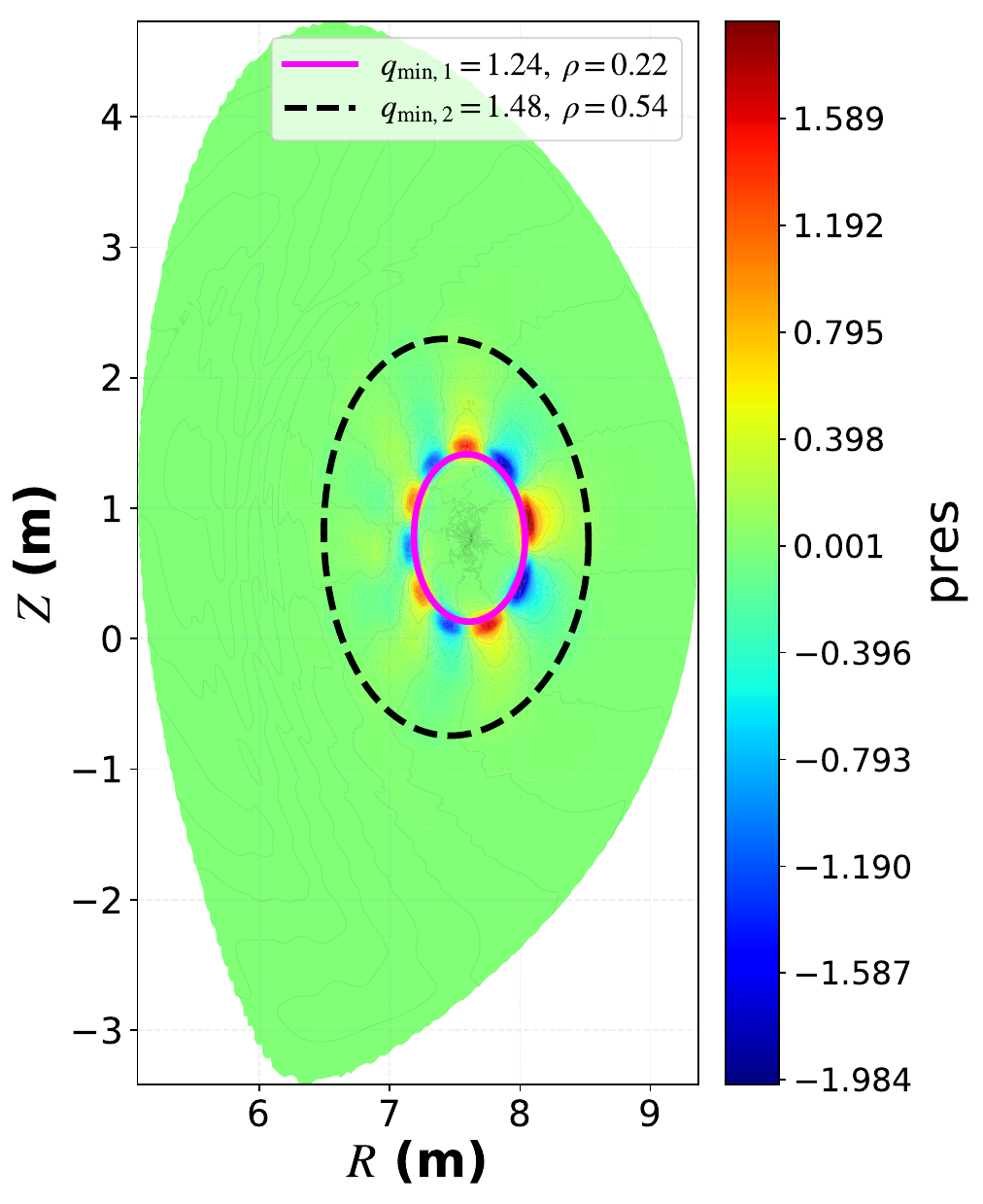}
        \label{fig:contour_q124_noep}
    }

    \caption{
    Two-dimensional contour plots of the perturbed pressure for representative low-frequency modes in absence of EPs. The four panels correspond to $q_{\min}=0.98$, $1.02$, $1.20$, and $1.24$, respectively. 
    }
    \label{fig:infernal_contours_noep}
\end{figure}
\clearpage

\begin{figure}[htbp]
    \centering
    \subfloat[$q_{\min}=1.02$]{
        \includegraphics[width=0.48\textwidth]{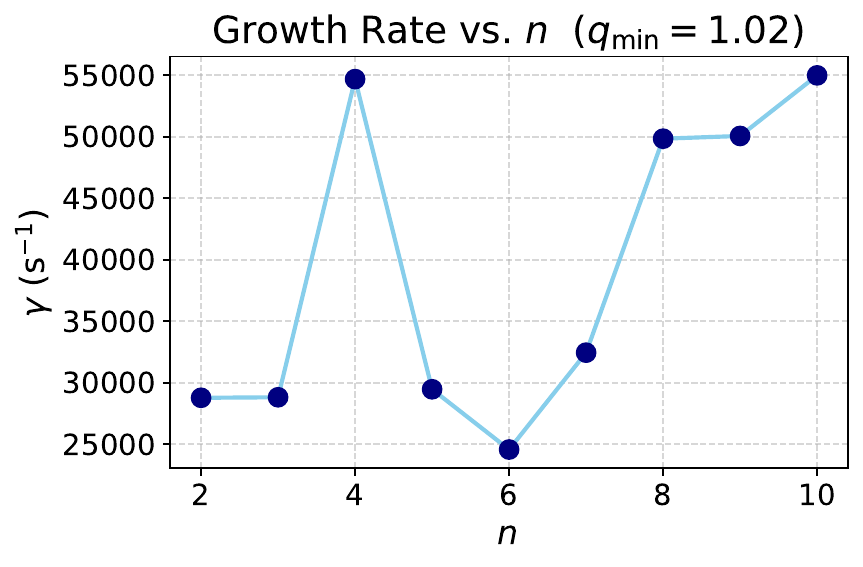}
        \label{fig:nscan_q102}
    }
    \subfloat[$q_{\min}=1.24$]{
        \includegraphics[width=0.48\textwidth]{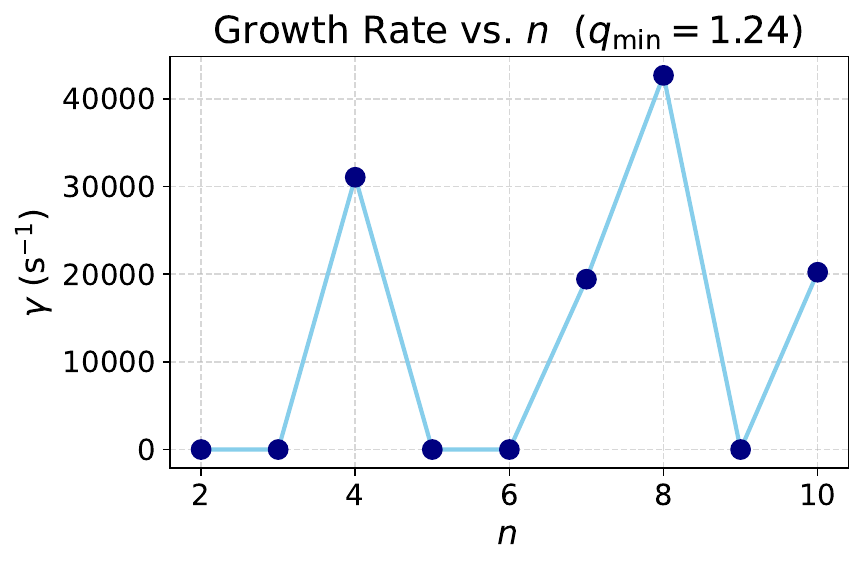}
        \label{fig:nscan_q124}
    }
    \caption{
    Growth rates of the infernal mode as functions of toroidal mode number $n$ for two representative cases: (a) $q_{\min}=1.02$ and (b) $q_{\min}=1.24$, both calculated without EPs.
}
    \label{fig:n_scan_infernal}
\end{figure}
\clearpage
\begin{figure}[htbp]
    \centering
    \includegraphics[width=0.78\textwidth]{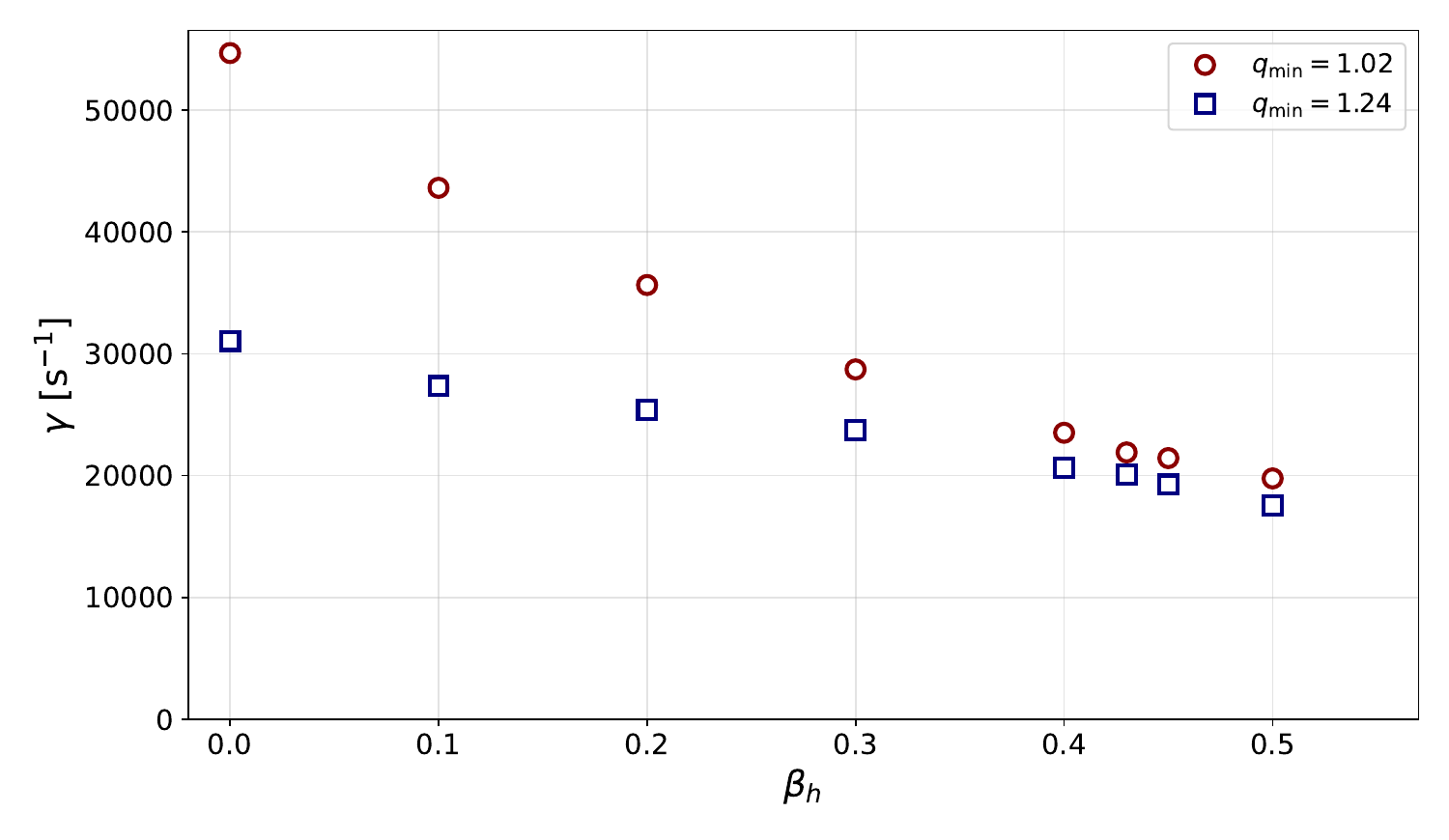}
    \caption{
    {Growth rates of representative infernal modes as functions of $\beta_h$ for $q_{\min}=1.02$ and $q_{\min}=1.24$.}}
    \label{fig:betah_scan_infernal}
\end{figure}
\clearpage
\begin{figure}[htbp] \centering \subfloat[ {Infernal mode, $q_{\min}=0.98$.} \label{fig:phase_space_ep_q098}] { \includegraphics[width=0.47\textwidth] {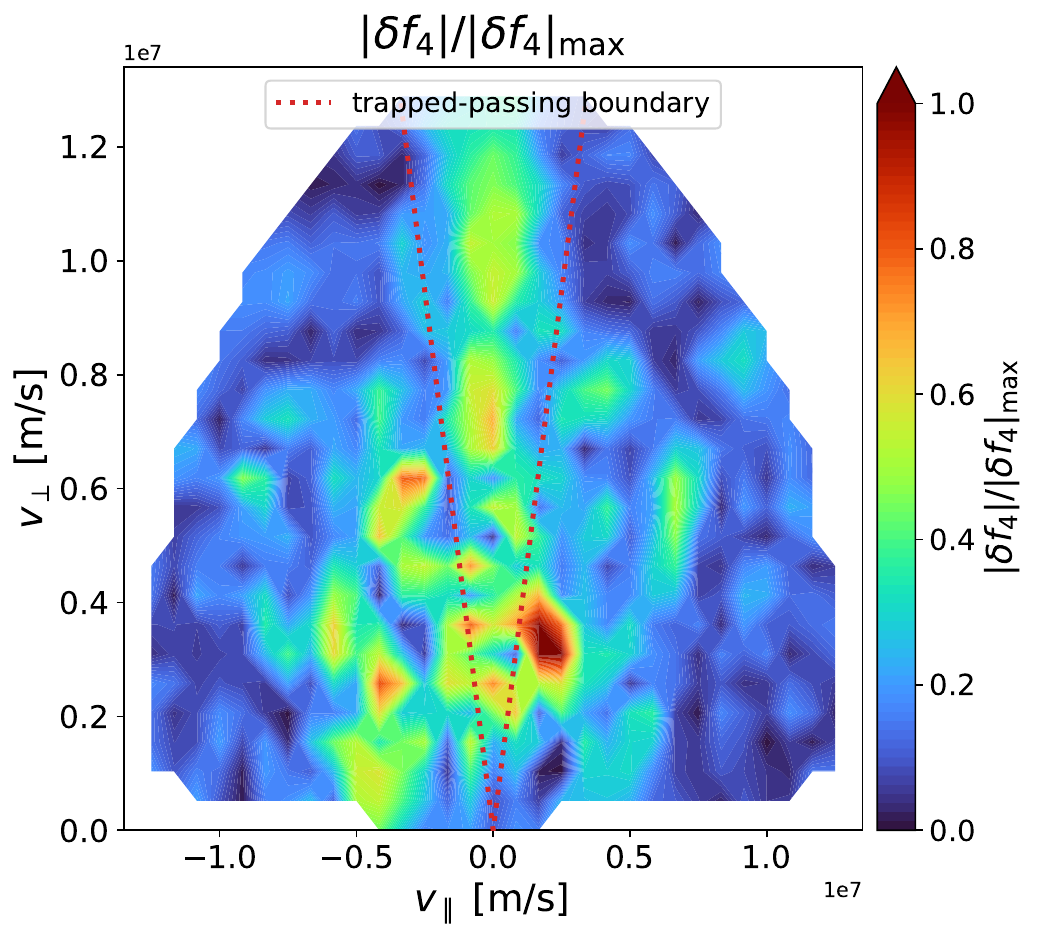} } \hfill \subfloat[ {EPM, $q_{\min}=1.10$.} \label{fig:phase_space_ep_q110}] { \includegraphics[width=0.47\textwidth] {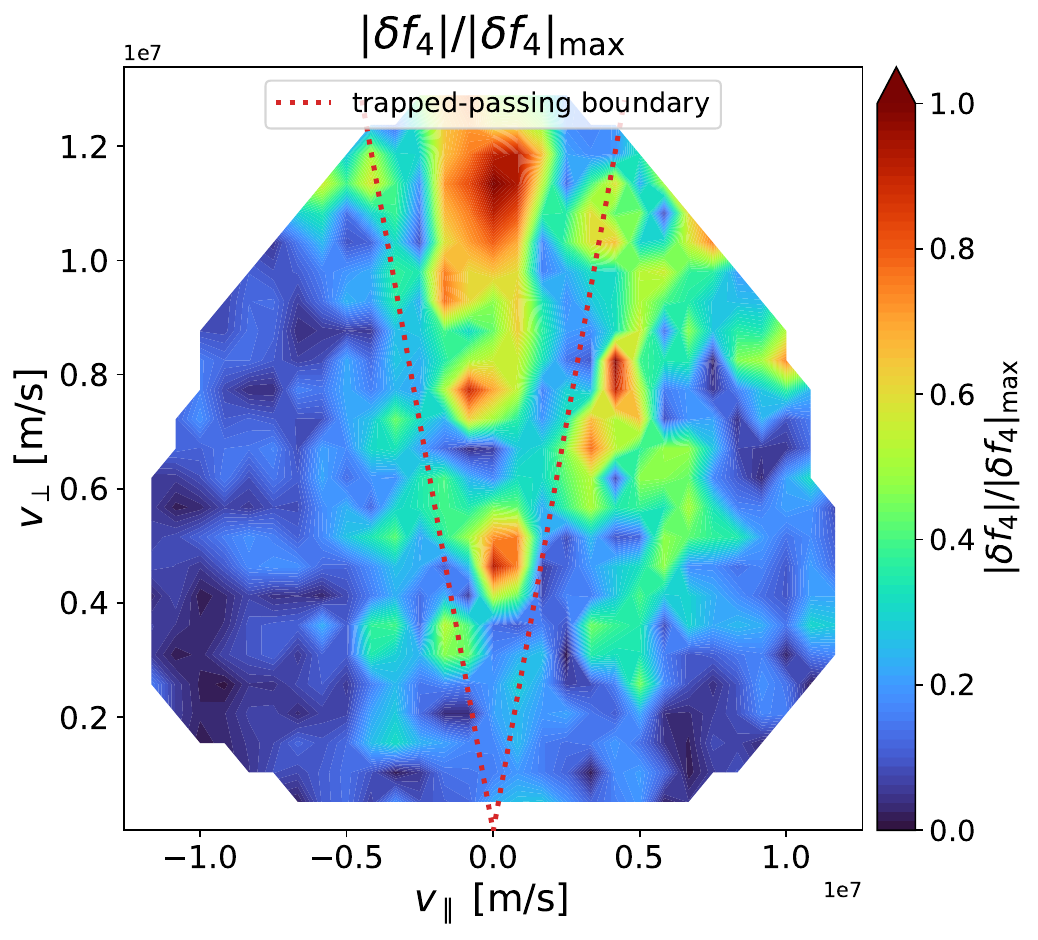} } \vspace{0.6em} \subfloat[ {Infernal mode, $q_{\min}=1.24$.} \label{fig:phase_space_ep_q124}] { \includegraphics[width=0.47\textwidth] {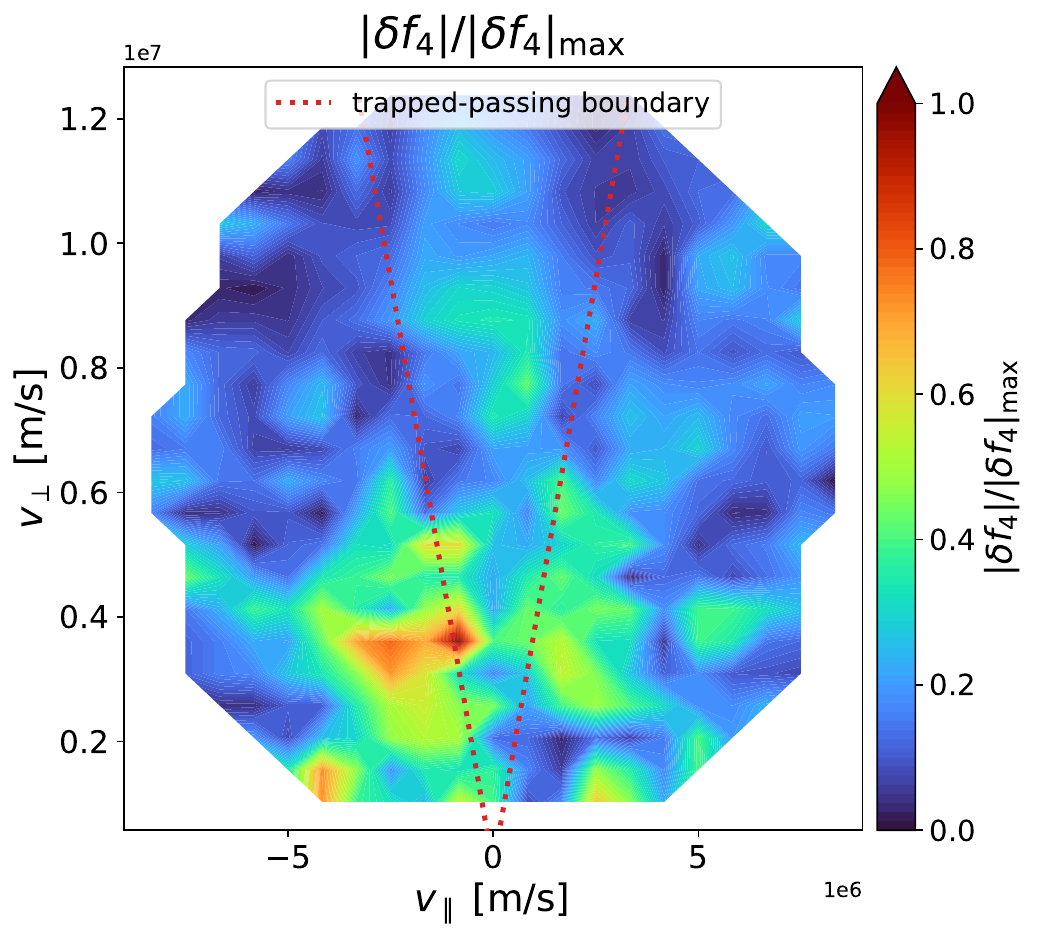} } \hfill \subfloat[ {RSAE, $q_{\min}=1.27$.} \label{fig:phase_space_ep_q127}] { \includegraphics[width=0.47\textwidth] {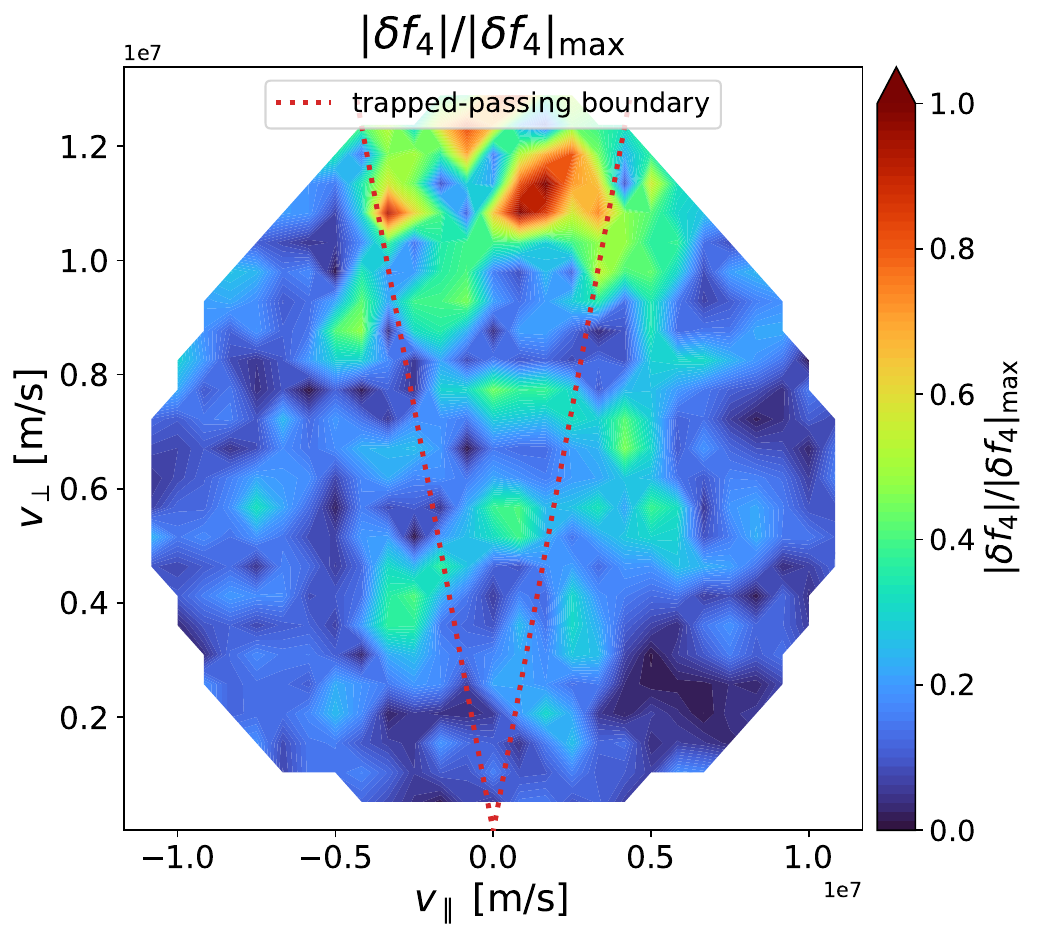} }

    \caption{
    Normalized EP responses $|\delta f_4|/|\delta f_4|_{\max}$ {in the phase-space ($v_\|,v_\perp$) plane} for representative mode branches{, where the dotted lines denote the corresponding trapped-passing boundaries.}}
    \label{fig:phase_space_ep}
\end{figure}

\clearpage

\begin{figure}[htbp] 
	\centering 
	\subfloat[ {Infernal mode, $q_{\min}=0.98$, $m=4$.} \label{fig:coherent_q098_m4}] { \includegraphics[width=0.47\textwidth] {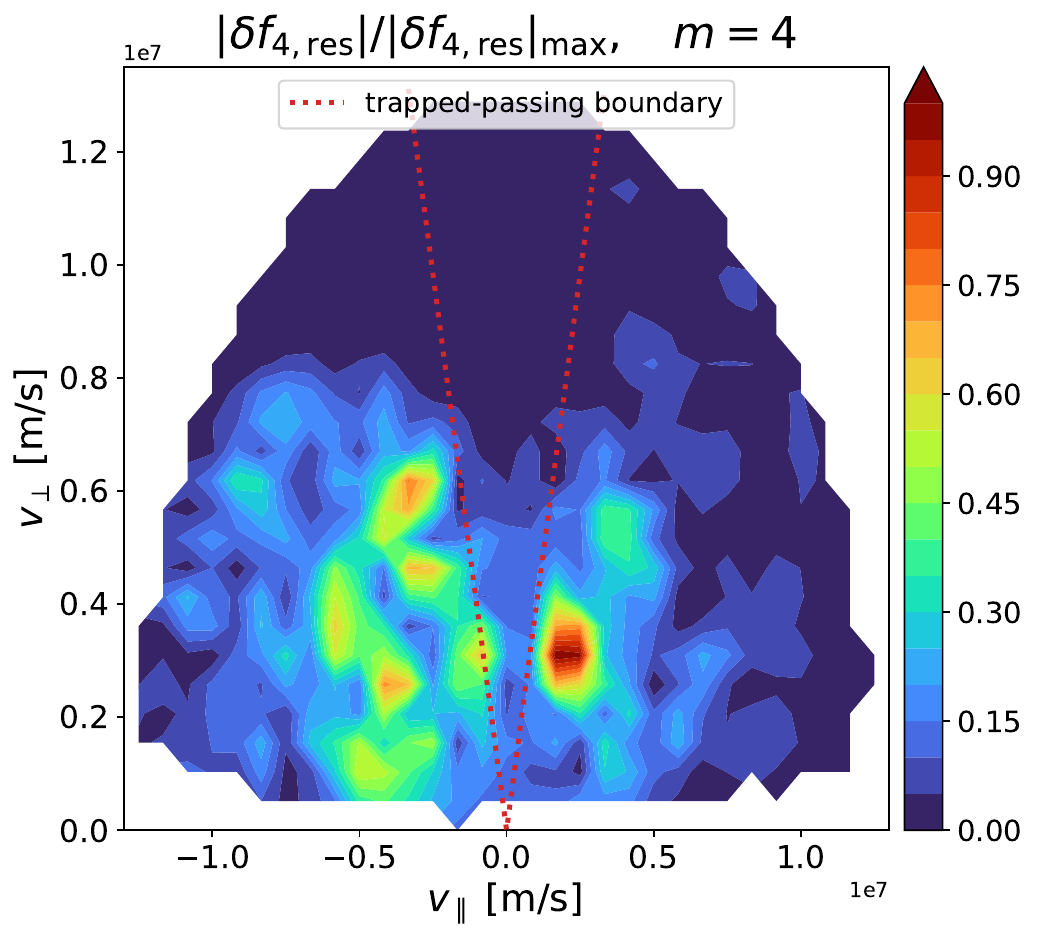} } \hfill \subfloat[ {EPM, $q_{\min}=1.10$, $l=-1$.} \label{fig:coherent_q110_lm1}] { \includegraphics[width=0.47\textwidth] {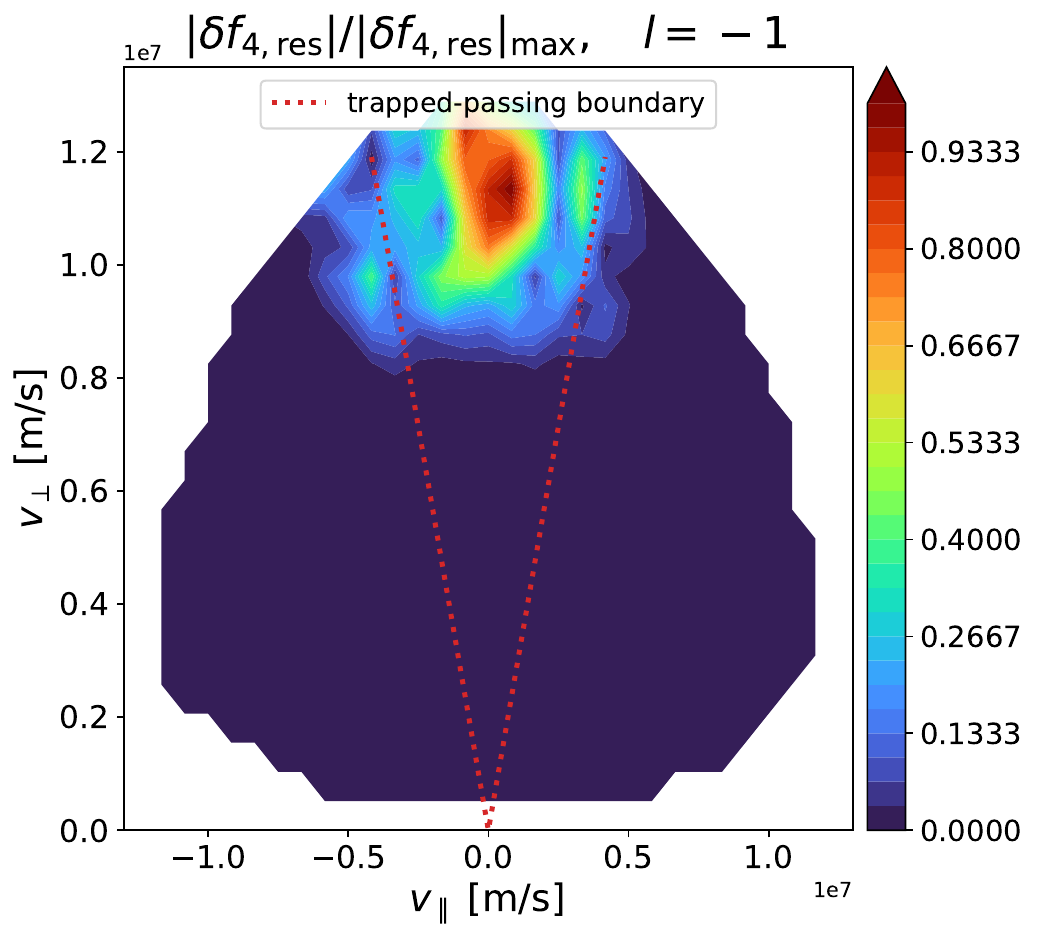} } \vspace{0.6em} \subfloat[ {Infernal mode, $q_{\min}=1.24$, $m=5$.} \label{fig:coherent_q124_m5}] { \includegraphics[width=0.47\textwidth] {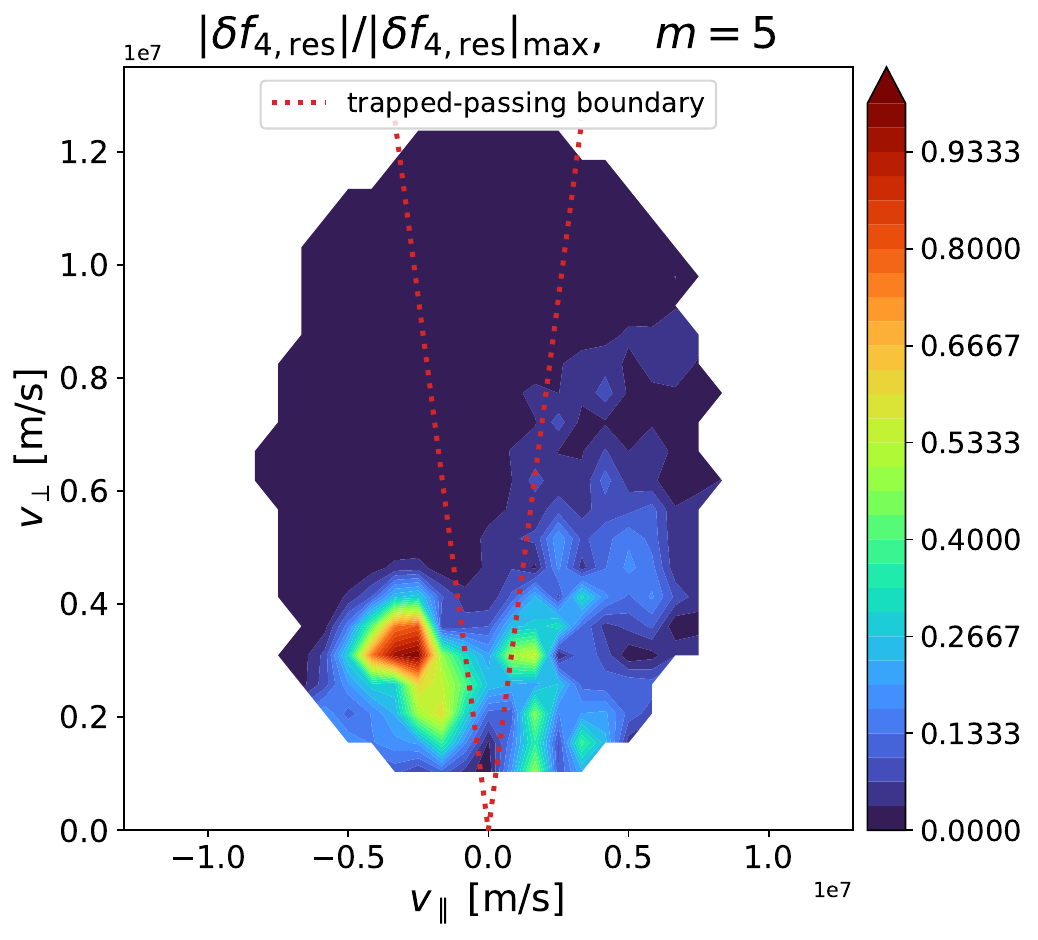} } \hfill \subfloat[ {RSAE, $q_{\min}=1.27$, $l=2$.} \label{fig:coherent_q127_l2}] { \includegraphics[width=0.47\textwidth] {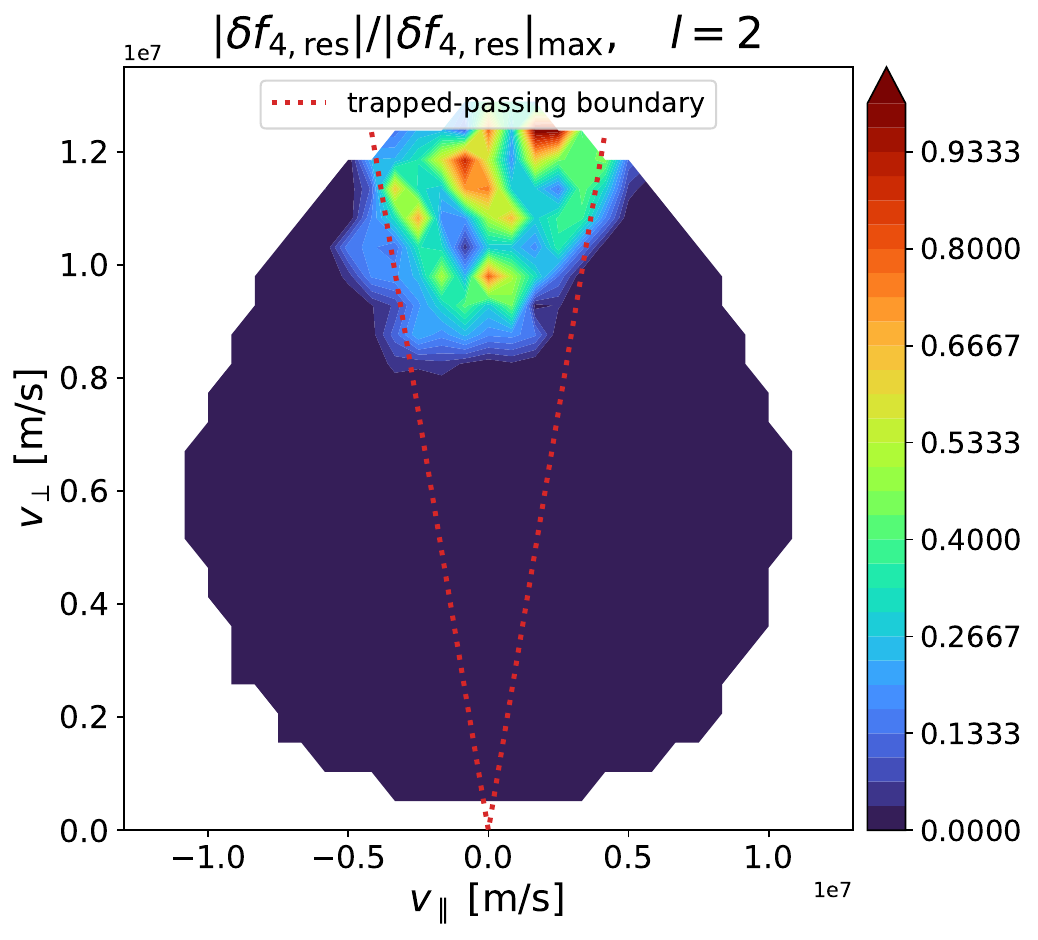} }

    \caption{ {The $n=4$ toroidal projections of resonant EPs in the {phase-space}
    $(v_\parallel,v_\perp)$ plane {satisfying} the indicated resonance conditions {in Eqs.~(\ref{eq:passing_resonance})--(\ref{eq:resonance_detuning}), where the dotted lines denote the corresponding trapped-passing boundaries.}}
    }
    \label{fig:coherent_contribution_ep}
\end{figure}
\clearpage
\begin{figure}[htbp] 
	\centering \subfloat[ { Infernal mode, $q_{\min}=0.98$, $\Lambda=0.34$. } \label{fig:pphi_energy_q098}] { \includegraphics[width=0.47\textwidth] {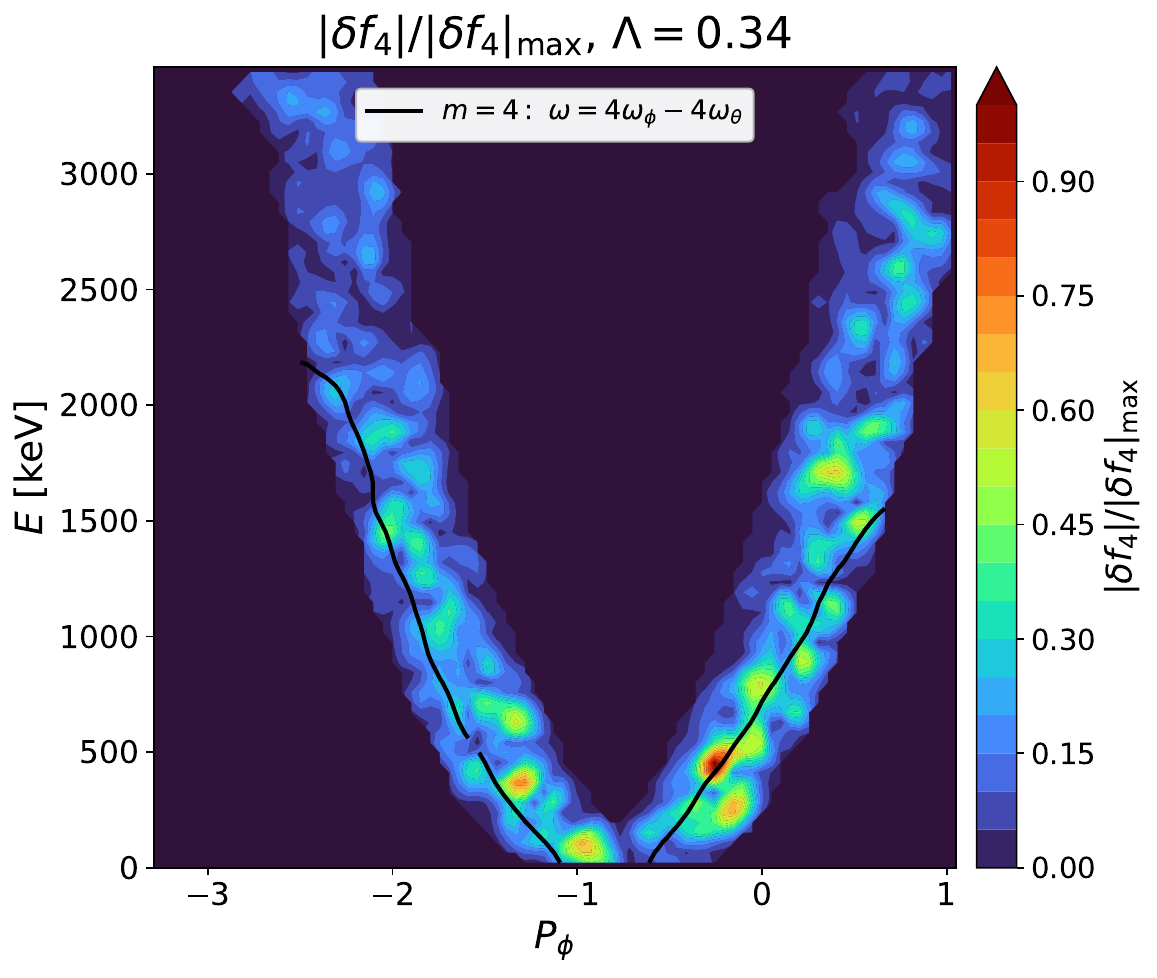} } \hfill \subfloat[ { EPM, $q_{\min}=1.10$, $\Lambda=1.08$. } \label{fig:pphi_energy_q110}] { \includegraphics[width=0.47\textwidth] {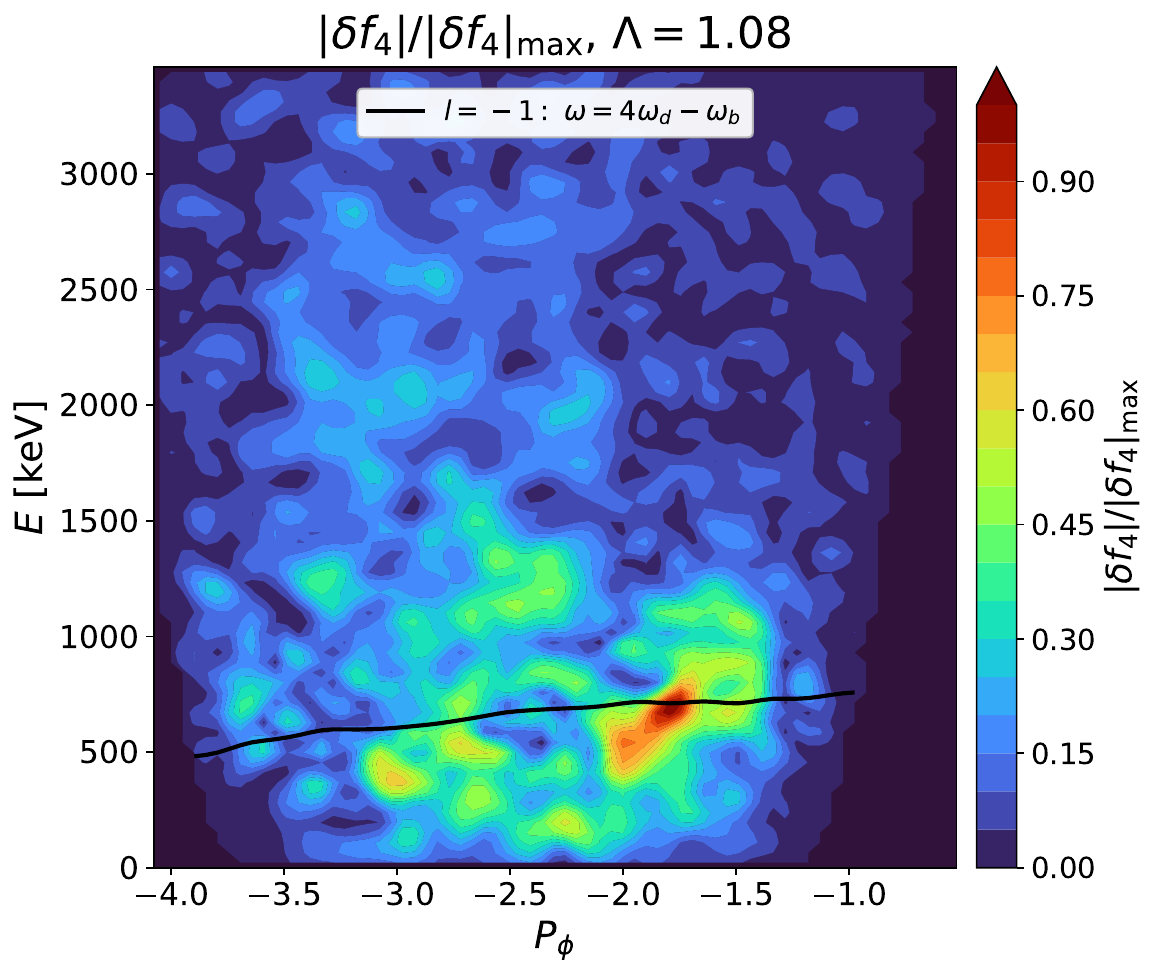} } \vspace{0.6em} \subfloat[ { Infernal mode, $q_{\min}=1.24$, $\Lambda=0.69$. } \label{fig:pphi_energy_q124}] { \includegraphics[width=0.47\textwidth] {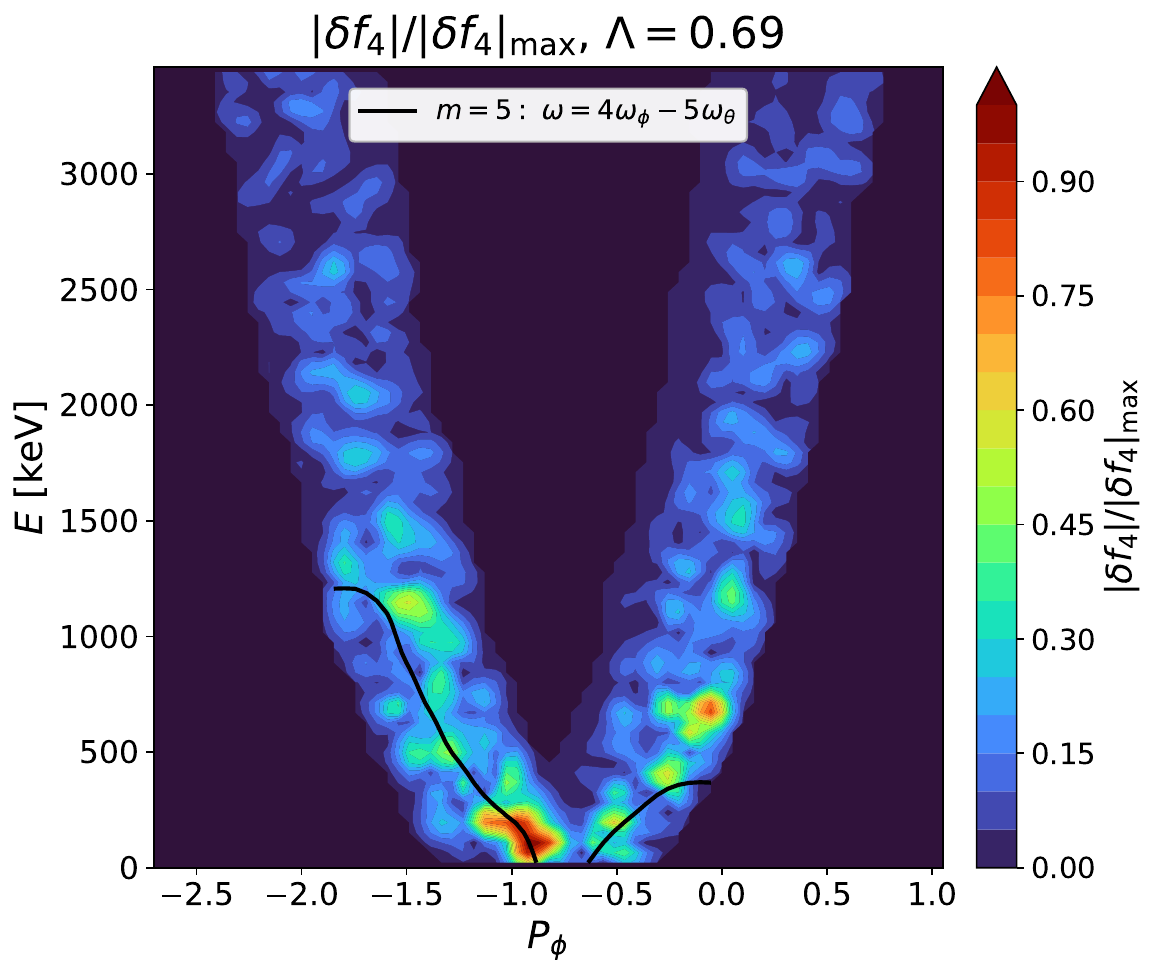} } \hfill \subfloat[ { RSAE, $q_{\min}=1.27$, $\Lambda=1.08$. } \label{fig:pphi_energy_q127}] { \includegraphics[width=0.47\textwidth] {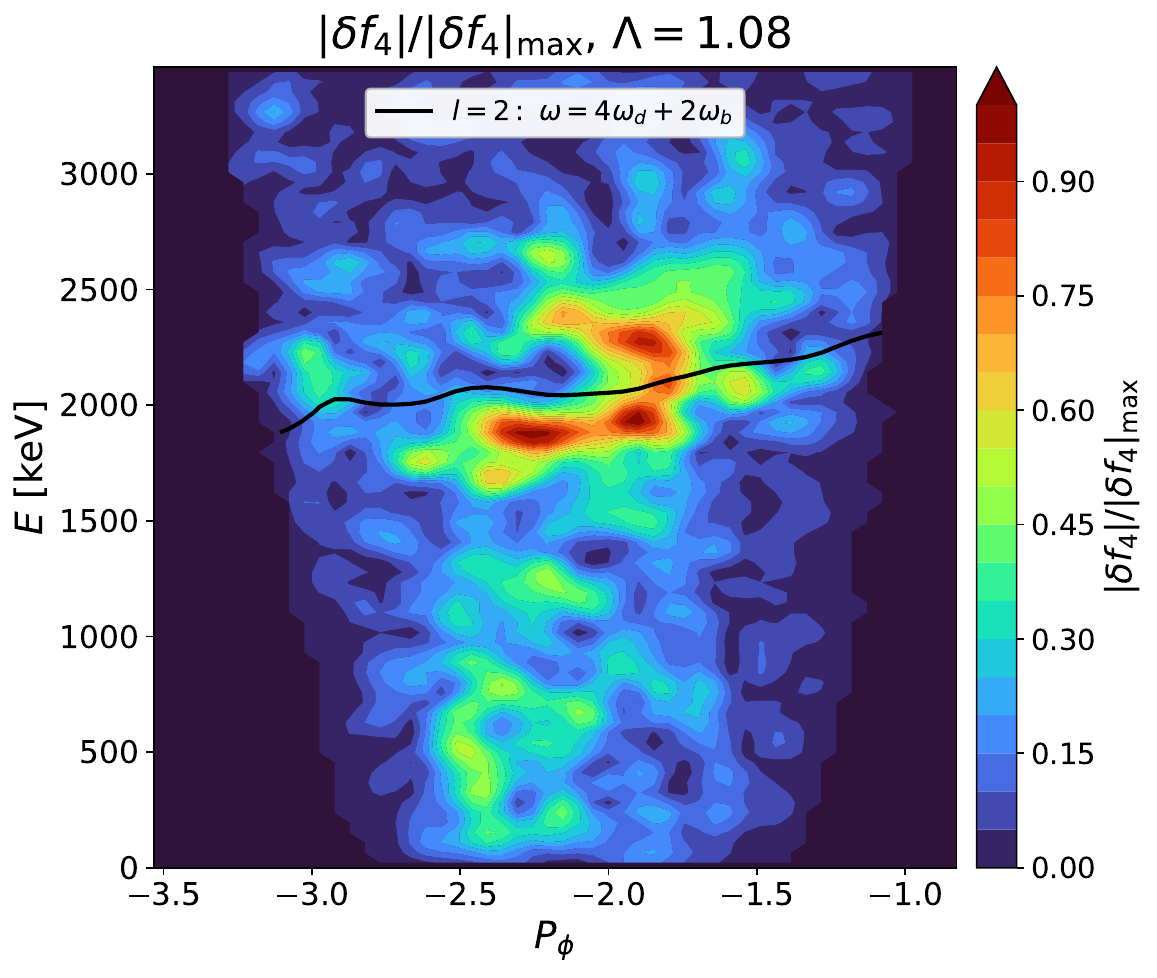} } 
	
	\caption{ { Normalized $n=4$ EP distribution perturbations in the $(P_\phi,E)$
			space for {various specific $\Lambda$ values of} pitch angles. The black curves denote the corresponding resonance lines {satisfying the resonance conditions in Eqs.~(\ref{eq:passing_resonance}) and (\ref{eq:trapped_resonance})}. }} \label{fig:pphi_energy_resonance} 
\end{figure}
\clearpage
\section*{References}

\bibliography{sample-1}

\end{document}